\documentclass[openacc]{rsproca_new}
\usepackage{epstopdf}

\begin{document}

\title{Generalised Quasilinear Approximation of the Interaction of Convection and Mean Flows in a Thermal Annulus }

\author{
S.M.Tobias$^{1}$, J. Oishi$^{2}$ and J.B. Marston$^{3}$}

\address{$^{1}$Department of Applied Mathematics, University of Leeds, Leeds, LS2 9JT, UK\\
$^{2}$Department of Physics
Bates College
Lewiston, ME 04240, USA\\
$^{3}$Department of Physics, Box 1843, Brown University, Providence, Rhode Island 02912-1843, USA}

\subject{applied mathematics, fluid mechanics, astrophysics}

\keywords{fluids mechanics, convection, zonal flow, quasilinear approximation}

\corres{S.M. Tobias\\
\email{S.M.Tobias@leeds.ac.uk}}

\begin{abstract}
In this paper we examine the interaction of convection, rotation and mean flows in a thermal annulus.
In this system mean flows are driven by correlations induced by rotation leading to non-trivial Reynolds stresses. The mean flows
act back on the convective turbulence acting as a barrier to transport. For this system we demonstrate that the Generalised Quasilinear
Approximation (GQL) \cite{mct2016} may provide a much better approximation to the complicated full nonlinear dynamics than
the widely used Quasilinear Approximation (QL). This result will enable the construction of more accurate statistical theories for 
the description of geophysical and astrophysical flows.
\end{abstract}


\begin{fmtext}

\section{Introduction}


A complete description of geophysical and astrophysical flows involves taking into account the nonlinear
interaction of fluids on a vast range of spatial and temporal scales. Such an undertaking via direct solution of the partial differential equations
(often termed Direct Numerical Simulation or DNS) is currently beyond the 
capability of numerical models, even utilising state-of-the art high performance computation on massively parallel
architectures \cite{bauer2015quiet,lorenz1967nature}.  A major problem for such schemes is that, for the rotating and stratified systems that are typical
of geophysical and astrophysical flows, the large scales influence and are in turn influenced by the smaller scales (which are
typically difficult to model)\cite{vallis:2016}.

\end{fmtext}
\maketitle

For this reason, much effort has been dedicated to deriving alternative approaches to DNS that in some way takes
into account the small-scale interactions in the large-scale dynamics (and the corresponding effect of large scales on the
small scale turbulence). One such approach is that termed Direct Statistical Simulation (DSS). In this approach
the statistics of the geophysical and astrophysical turbulence are solved for (rather than the detailed dynamics) \cite{tobias2011astrophysical}. In principle,
numerical solution of Fokker-Planck equations may lead to the determination of the pdf of the statistics of turbulent flows, or large deviation theory 
may give indications of the probabilities of rare events. One such example of DSS that has received much attention recently
is a generalisation of the methods of Kolomogorov and Kraichnan (see e.g. \cite{frisch1995turbulence}) to flows that are inhomogeneous and anisotropic (as is typically
the case for geophysical and astrophysical flows --- owing to the presence of mean flows, rotation, stratification and possibly magnetic fields.
This (equal-time) cumulant expansion method of DSS is able to capture the large scales of turbulent flows
with fewer degrees of freedom, as the low-order statistics are
spatially smoother than the corresponding dynamical fields and are described by evolution on a slow manifold \cite{marston2014direct}.

In order to yield tractable computational problems, the cumulant expansion should be truncated as soon as
possible. Truncated at second order (CE2) the method has been shown adequately to represent the
statistics of planetary jets and those of the nonlinear magnetorotational instability \cite{sb2015}. Furthermore at this level
of truncation the method can be shown to be formally equivalent to  the stochastic
structural stability theory (SSST or S3T) of Farrell, Ioannou and collaborators \cite{Farrell:2007fq,Constantinou:2013fh}. This method
has been utilised to model a number of physical systems and can be
justified for systems near equilibrium for which there is a separation of time-scales
\cite{Bouchet:2018er}. At this level of truncation the cumulant expansion is the the statistical
representation of the Quasilinear (QL) approximation (see e.g. \cite{marston65conover}). However it can be shown that for systems
far from statistical equilibrium that CE2 based on zonal averaging (and hence QL) ceases to be an accurate representation \cite{tobias2013direct}.
For such systems there are three possibilities for deriving a system that yields a more  accurate representation of the statistics
of out-of-equilibrium systems. The first is to include eddy/eddy $\rightarrow$ eddy
 in the cumulant expansion truncation. This leads to the CE3 (or CE2.5) approximation, which
has been shown to improve performance of DSS   \cite{marston2014direct}. A second possibility is to extend the averaging procedure from zonal averaging to 
ensemble averaging (see e.g. \cite{bakas2013emergence,atm2017}), thus enabling the description of coherent structures other than
zonal means. Finally one may
generalise the quasilinear approximation \cite{mct2016}, as described below, and derive a corresponding statistical theory (GCE2).

The Generalised Quasilinear Approximation (GQL approximation) was introduced by Marston et al (2016) and its
effectiveness in reproducing the fully nonlinear results has been compared with that of the QL approximation
for a number of systems including the driving of barotropic jets, the helical magnetorotational instability (HMRI) and
rotating couette flow \cite{mct2016,chmt2016,tm2017}. In all of these systems the driving of the velocity arises either via a small-scale forcing
or through the boundaries of the system and is independent of the dynamics of the system. Another class of problems, 
important for geophysical and astrophysical flows, exists however. In this class of convective systems the driving arises 
through buoyancy and the effectiveness of the driving input into kinetic energy may depend on the state of the system.
In this paper we examine the simplest geophysically realistic system that involves the interaction of rotating convection
with mean (zonal) flows --- the Busse annulus model. We shall determine the effectiveness of the QL and GQL approximations
for this system which may undergo many types of non-trivial dynamics.

In the next section we give a brief derivation of the model and the equations, before showing some representative dynamics
in section 3. In Section 4 we consider the QL and GQL approximations before giving our conclusions in the final section.

\section{Model and Equations}

\begin{figure}[!h]
\centering
\includegraphics[width=4.5in]{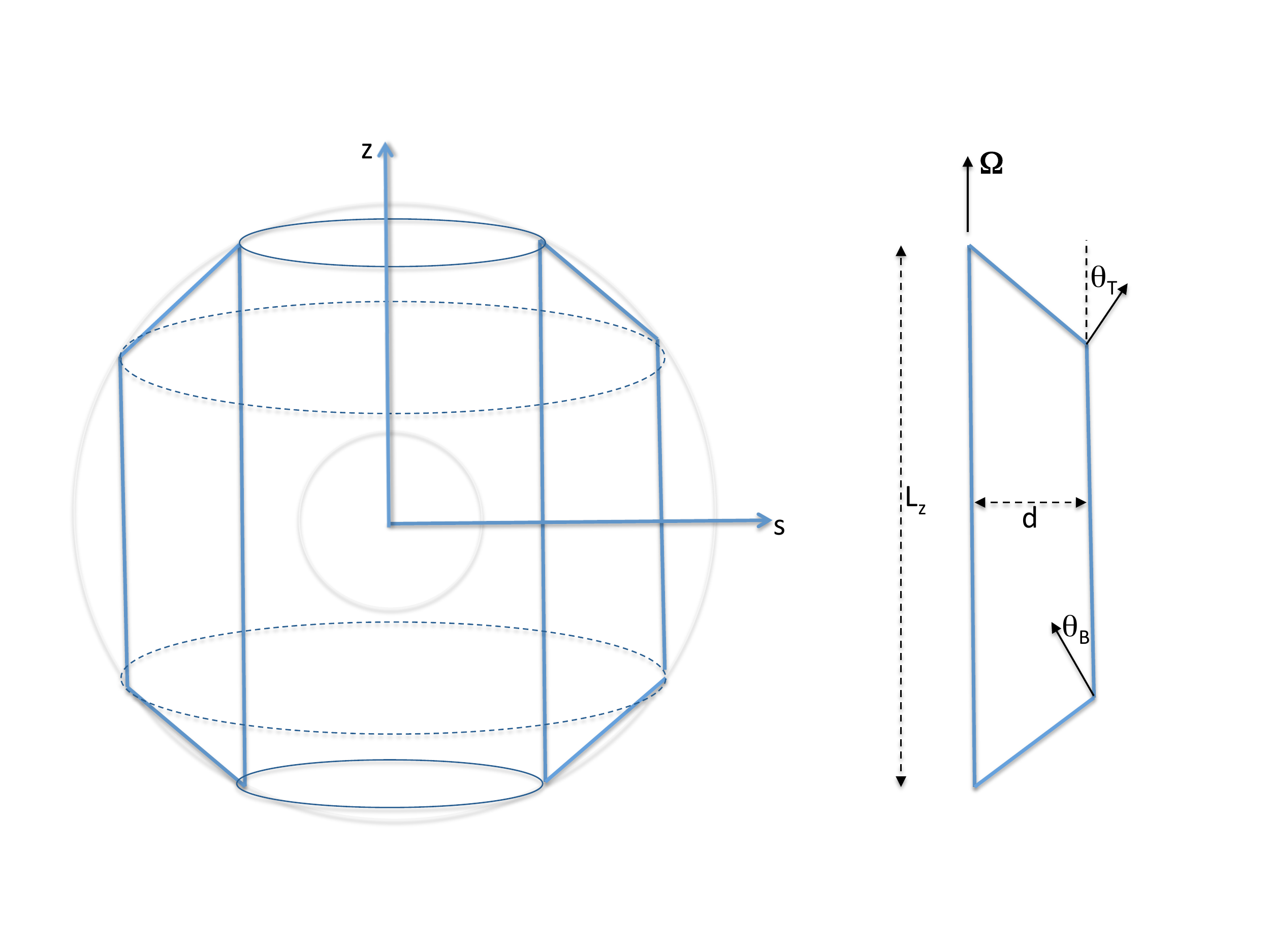}
\caption{Schematic diagram of computational domain for the thermal annulus..}
\label{fig_busse_pic}
\end{figure}

The model we consider here is of rotating incompressible Boussinesq fluid (with viscosity $\nu$ and thermal diffusivity $\kappa$) in an annulus of length $L_x$ with vertical sidewalls and weakly tilted top and bottom boundaries (see figure~\ref{fig_busse_pic} and  \cite{busse:1976,bh1993,rj2006}). The system rotates at angular velocity ${\boldsymbol \Omega = \Omega {\bf e}_z}$ and gravity is uniform and in the $y$-direction. The angle of inclination of the top and bottom boundaries are given by $\theta_T$ and $\theta_B$ respectively. 
The system is non-dimensionalised, scaling lengths with the width of the annulus in the $y$-direction ($d$), time with the viscous timescale $d^2/\nu$ velocites with $\nu/d$ and temperatures with the temperature difference between the inner and outer walls ($\Delta T$). The dynamics of the system is then controlled by evolution equations for the non-dimensional velocity ${\bf u}$ and temperature $T$, which depend on three non-dimensional parameters; the modified Rayleigh number, the Ekman number and the Prandtl number given by
\begin{equation}
Ra^\prime = \dfrac{\alpha g \Delta T d}{\nu \Omega}, \quad E = \dfrac{\nu}{\Omega d^2}, \quad Pr = \dfrac{\nu}{\kappa}.
\end{equation}
Here $\alpha$ is the coefficient of thermal expansion at constant pressure.

The temperature is decomposed $T = T_{BS} +{\hat \theta(x,y)}$ into a basic state profile $T_{BS}$ satisfying $\nabla^2 T_{BS} = 0$ and a perturbation ${\hat \theta}$. Here we take $T_{BS} = y$ for the basic state. Once stress-free boundary conditions have been imposed on the vertical walls and no-slip boundary conditions are imposed on the top and bottom boundaries then for two-dimensional quasi-geostrophic models 
the velocity and temperature can be written as
\begin{equation}
{\bf u} = -\boldsymbol{\nabla} \times  (\psi(x,y) {\bf e}_z) + {\bf u}^\prime({\bf r})
\end{equation}
with 
\begin{equation}
| {\bf u}^\prime| \ll | \boldsymbol{\nabla} \times  (\psi(x,y) {\bf e}_z)|
\end{equation}
and 
\begin{equation}
{\hat \theta} = {\hat \theta}(x,y),
\end{equation}
and the evolution equations for $\psi$ and ${\hat \theta}$ are given by (see \cite{rj2006})
\begin{eqnarray}
\dfrac{\partial \nabla^2 \psi}{\partial t} + J(\psi, \nabla^2 \psi) - \beta \dfrac{\partial \psi}{\partial x} &=& -\dfrac{Ra}{Pr} \, \dfrac{\partial {\hat \theta}}{\partial x} - C |\beta|^{1/2} \nabla^2 \psi + \nabla^2 \nabla^2 \psi,
\label{psieqn}\\
\dfrac{\partial {\hat \theta}}{\partial t} + J(\psi, {\hat \theta}) &=& - \dfrac{\partial {\psi}}{\partial x} + \dfrac{1}{Pr} \nabla^2 {\hat \theta},
\label{psiTeqns}
\end{eqnarray}
where $J(A,B) = \frac{\partial A}{ \partial x} \frac{\partial B}{ \partial y} - \frac{\partial A}{ \partial y} \frac{\partial B}{ \partial x}$ is the Jacobian and the parameters are given by \begin{equation}
Ra = \dfrac{\alpha g \Delta T d^3}{ \nu \kappa}, \quad \beta = \dfrac{(2 (\theta_T - \theta_B)d)}{(L_z E)}, \quad C = \left(\dfrac{2d}{|\theta_T-\theta_B| L_z}\right)^{1/2}\quad Pr = \dfrac{\nu}{\kappa}.
\end{equation}
Hence $\beta$ measures the degree of vortex stretching engendered by the sloping endwalls and $C$ measures the degree of friction. In general increasing $\beta$ and $C$ is expected to lead to an increase in the number of the jets (see later).

It is equations~(\ref{psieqn}-\ref{psiTeqns}) and their QL and GQL counterparts (given in section~\ref{GQL}) that form the basis of this paper. These are integrated numerically using the pseudospectral PDE solving package Dedalus (REF)

\subsection{Representative Dynamics.}

The system described above has been extensively studied utilising DNS by a number of authors (for example \cite{busse:1976,bh1993,rj2006}). The dynamics of the system is complicated and involves the interaction of thermal convection, rotation and zonal flows. The dynamics is known to depend critically on the rotation rate (as measured by the Ekman number and hence $\beta$) and the degree of supercriticality (as measured by the Rayleigh number). Less is known about the role of the Prandtl number in the system, though in related systems this has been shown to have a significant role in determining the form of the solutions \cite{gc2016}. In this paper we fix $Pr=1$ and and show representative dynamics for varying $\beta$, $C$ and Rayleigh number $Ra$.

\begin{center}
\begin{table}[!h]
\caption{Parameters for DNS runs}
\label{table}
\begin{tabular}{cccccc}
\hline
Run &$\beta$ &$Ra$ &$C$&Solution & Resolution ($ny \times nz$) \\
\hline
A &$2.8 \times 10^3$ &$7.6 \times 10^4$ &$0$ &Large-scale jets& $256 \times 64$\\
B &$7.07 \times 10^5$ &$1 \times 10^8$ & $0.316$&Multiple jets& $512 \times 256$ \\
C &$5.00 \times 10^5$  & $8 \times 10^7$& $0$&Bursting Jets&  $512 \times 256$\\\hline
\end{tabular}
\vspace*{-4pt}
\end{table}
\end{center}

\begin{figure}[!h]

\centering
\includegraphics[width=2.5in]{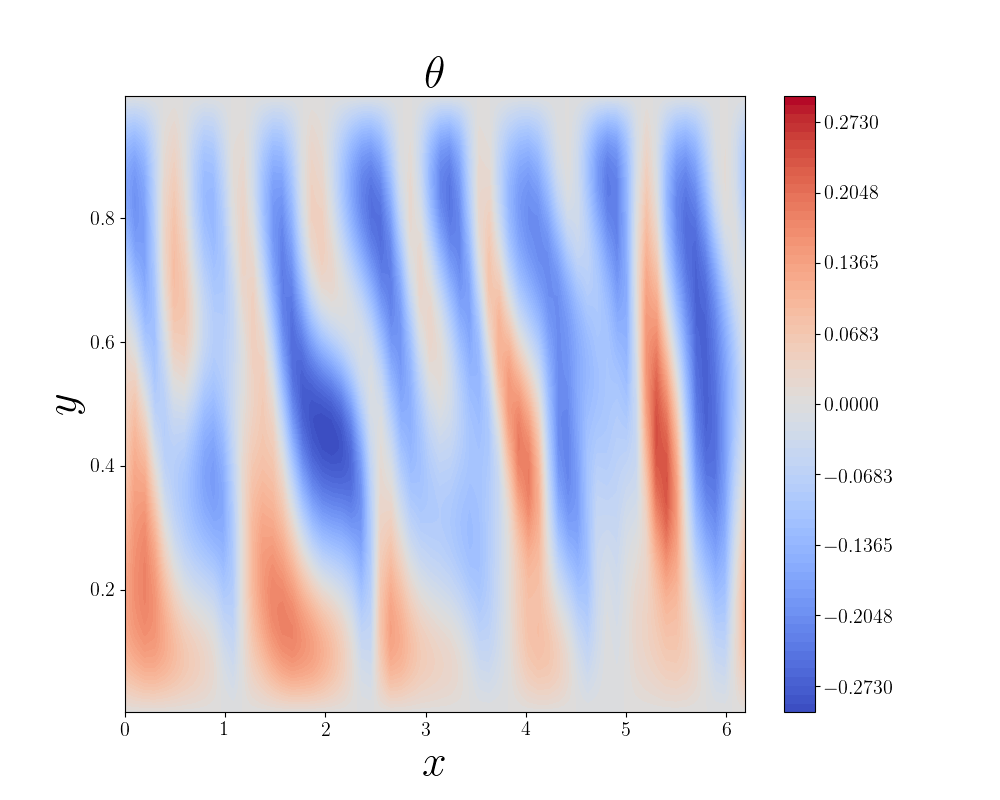}
\includegraphics[width=2.5in]{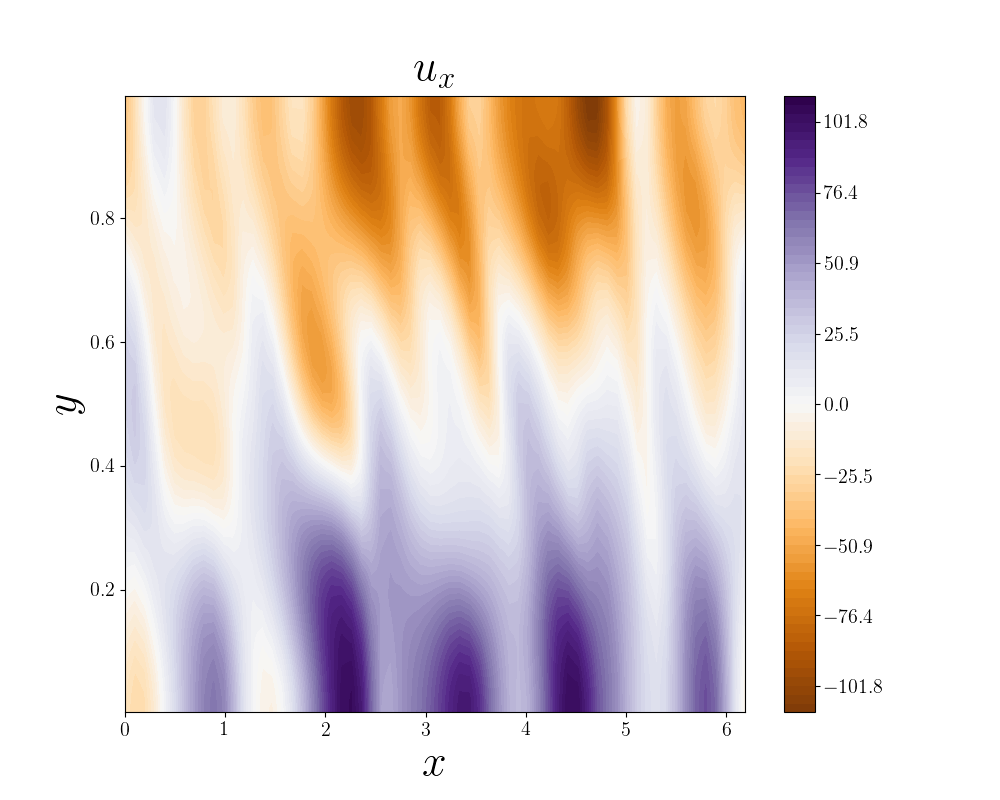}
\includegraphics[width=2.5in]{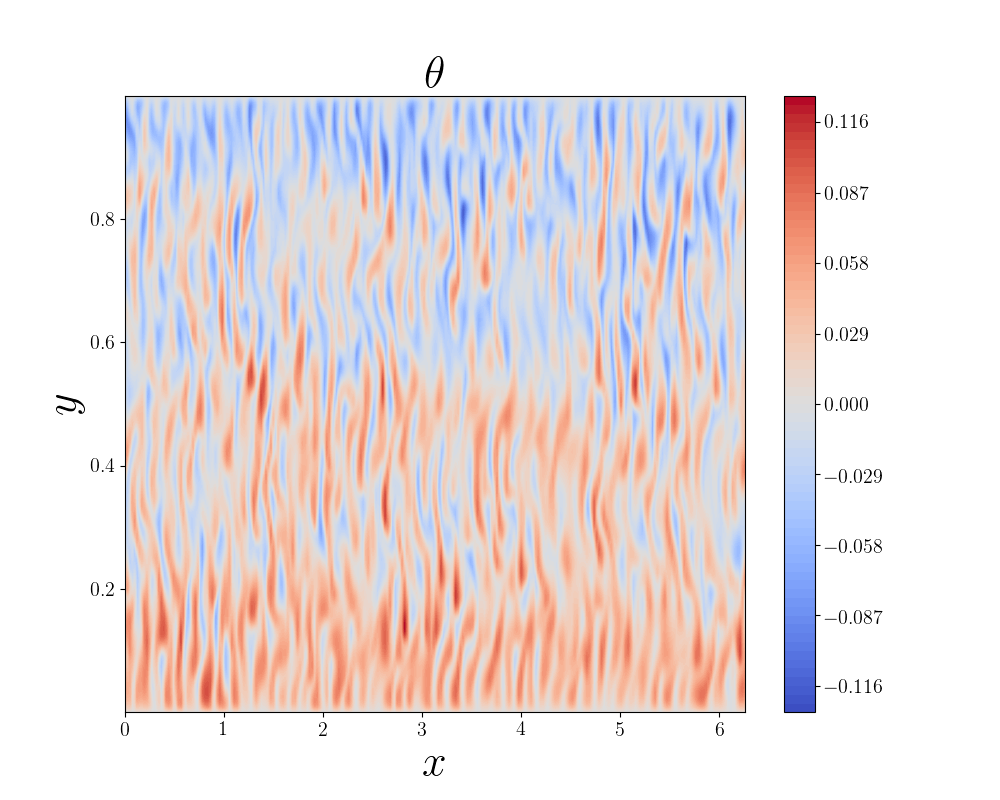}
\includegraphics[width=2.5in]{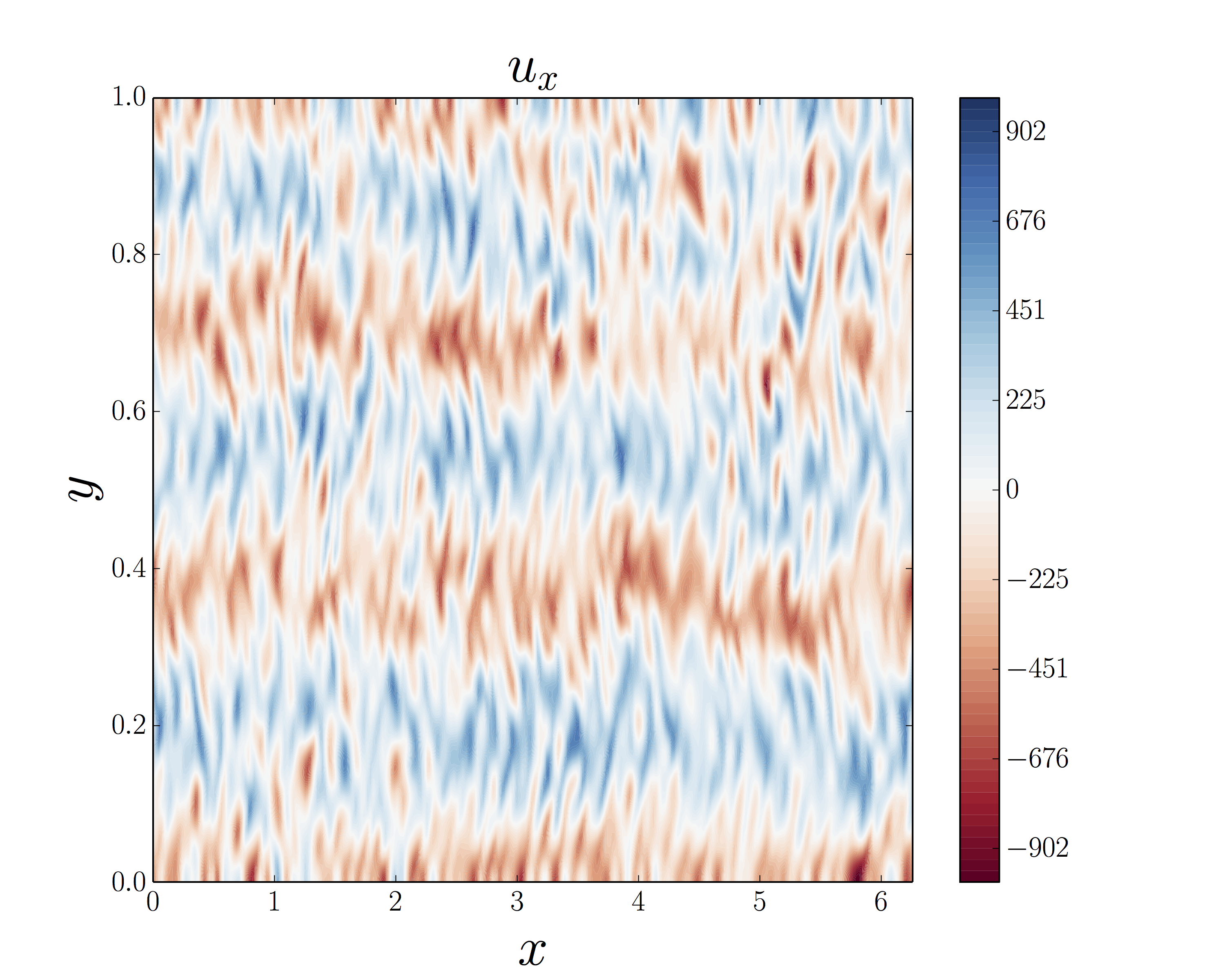}
\includegraphics[width=2.5in]{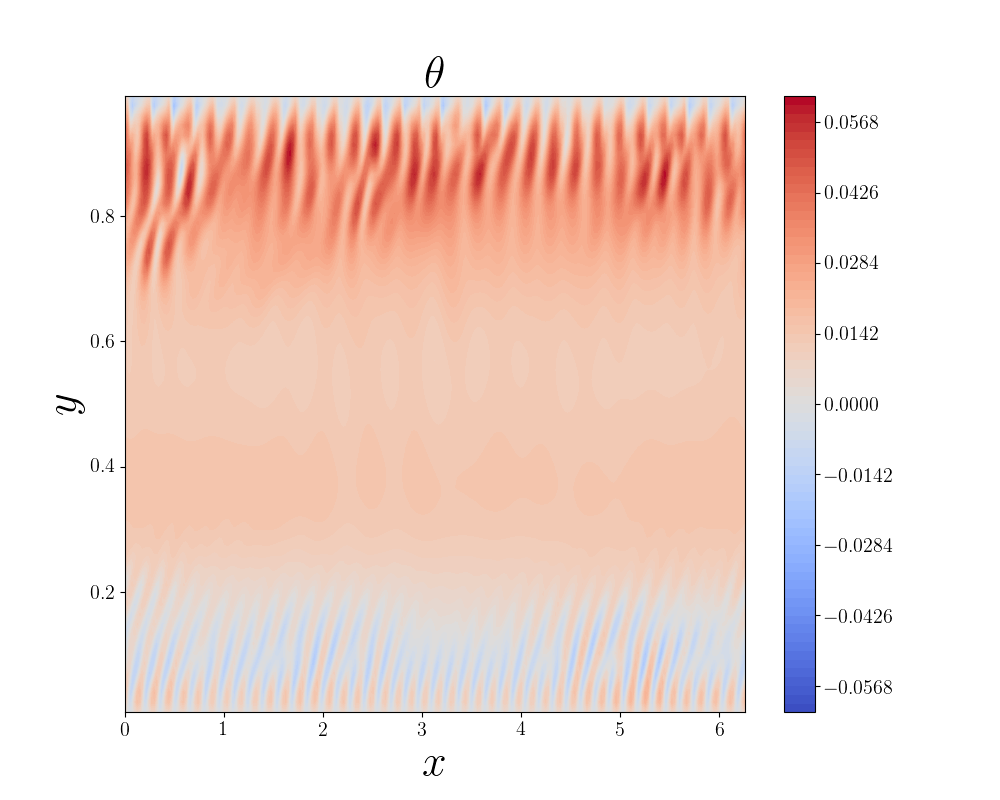}
\includegraphics[width=2.5in]{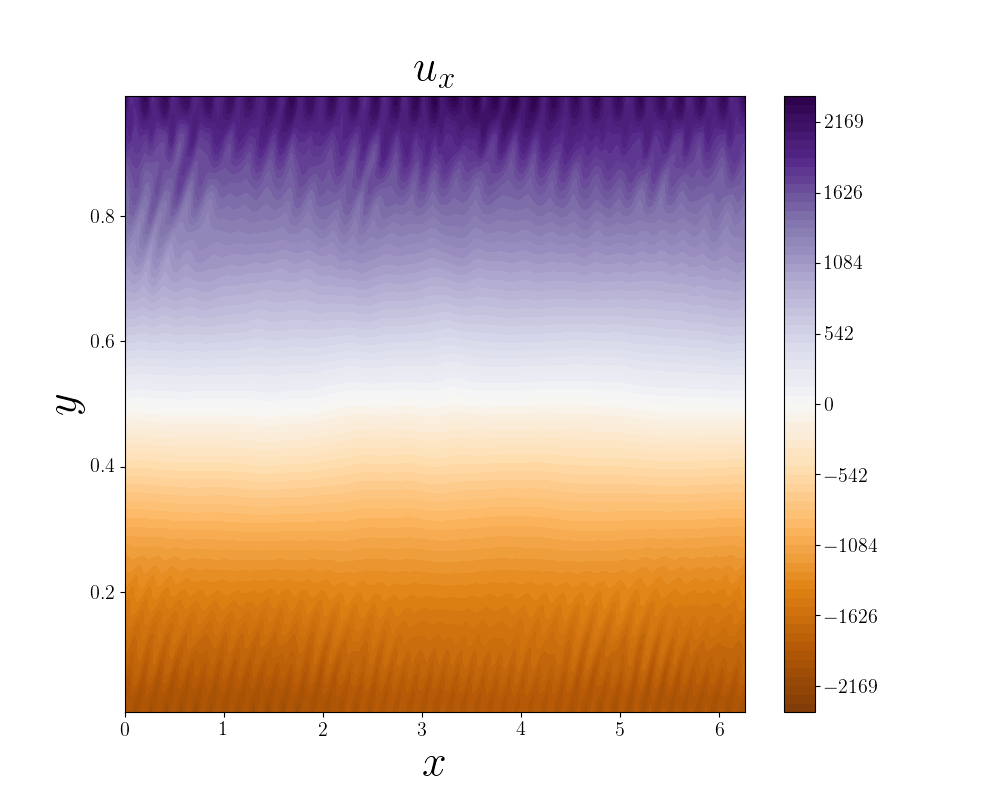}
\includegraphics[width=2.5in]{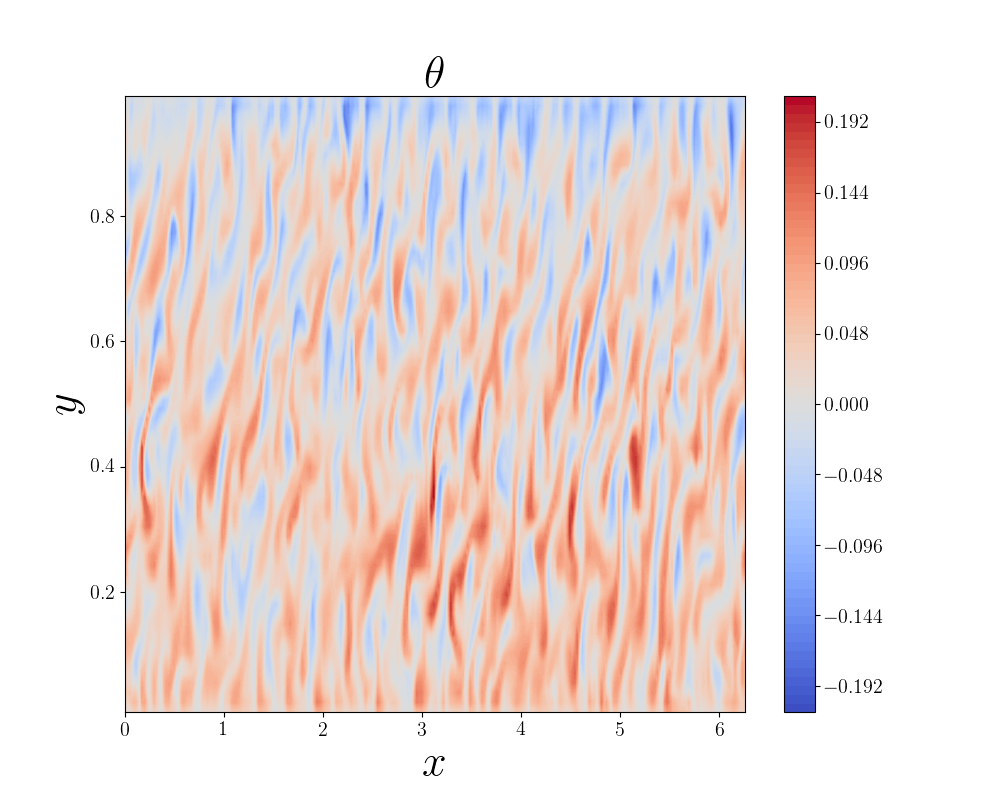}
\includegraphics[width=2.5in]{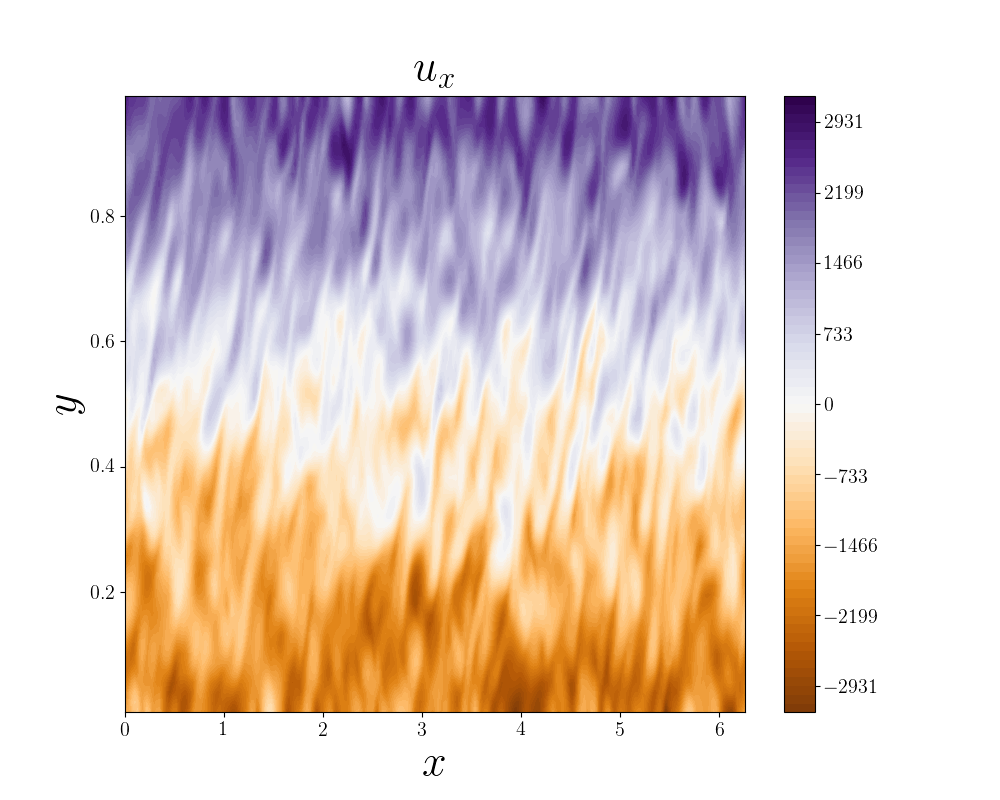}
\caption{Snapshots of horizontal velocity $u_x$ and $\theta$ in the saturated regime for (a,b) Case A (c,d) Case B (e,f) CASE C (during the strong shear phase) (d) CASE C (during the weak shear phase).}
\label{fig_dyn_snap}
\end{figure}

\begin{figure}[!h]
\centering\includegraphics[width=2.5in, height=2in]{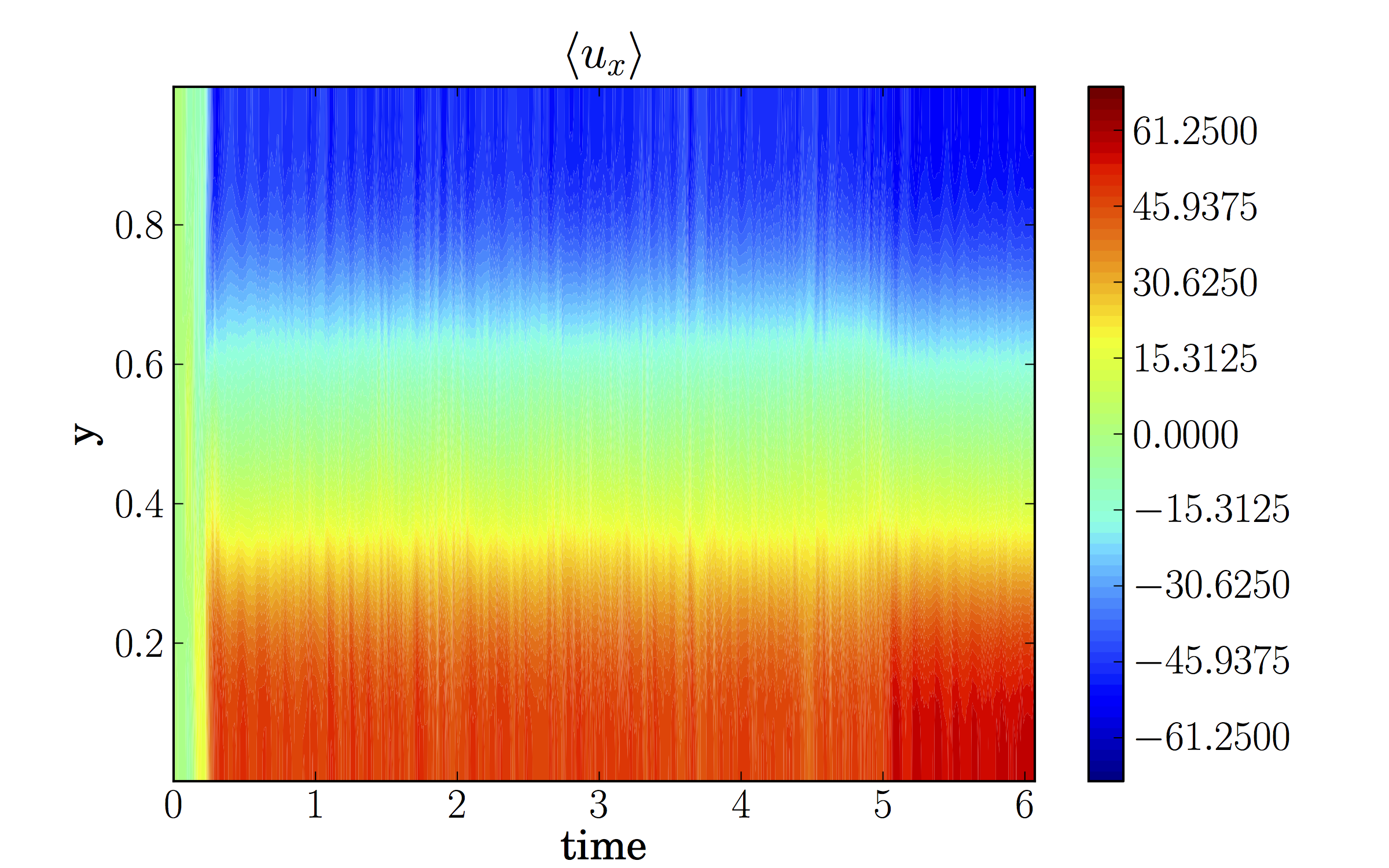}
\includegraphics[width=2.5in, height=2in]{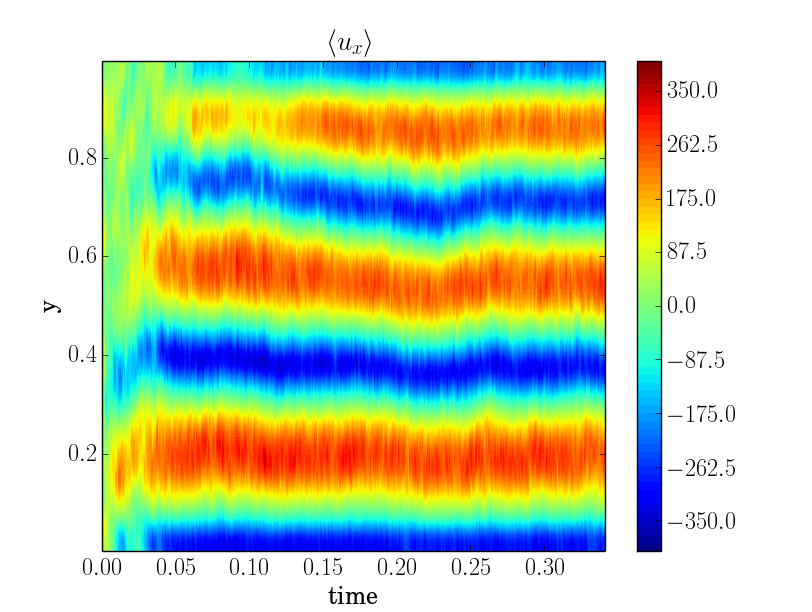}
\includegraphics[width=2.5in, height=2in]{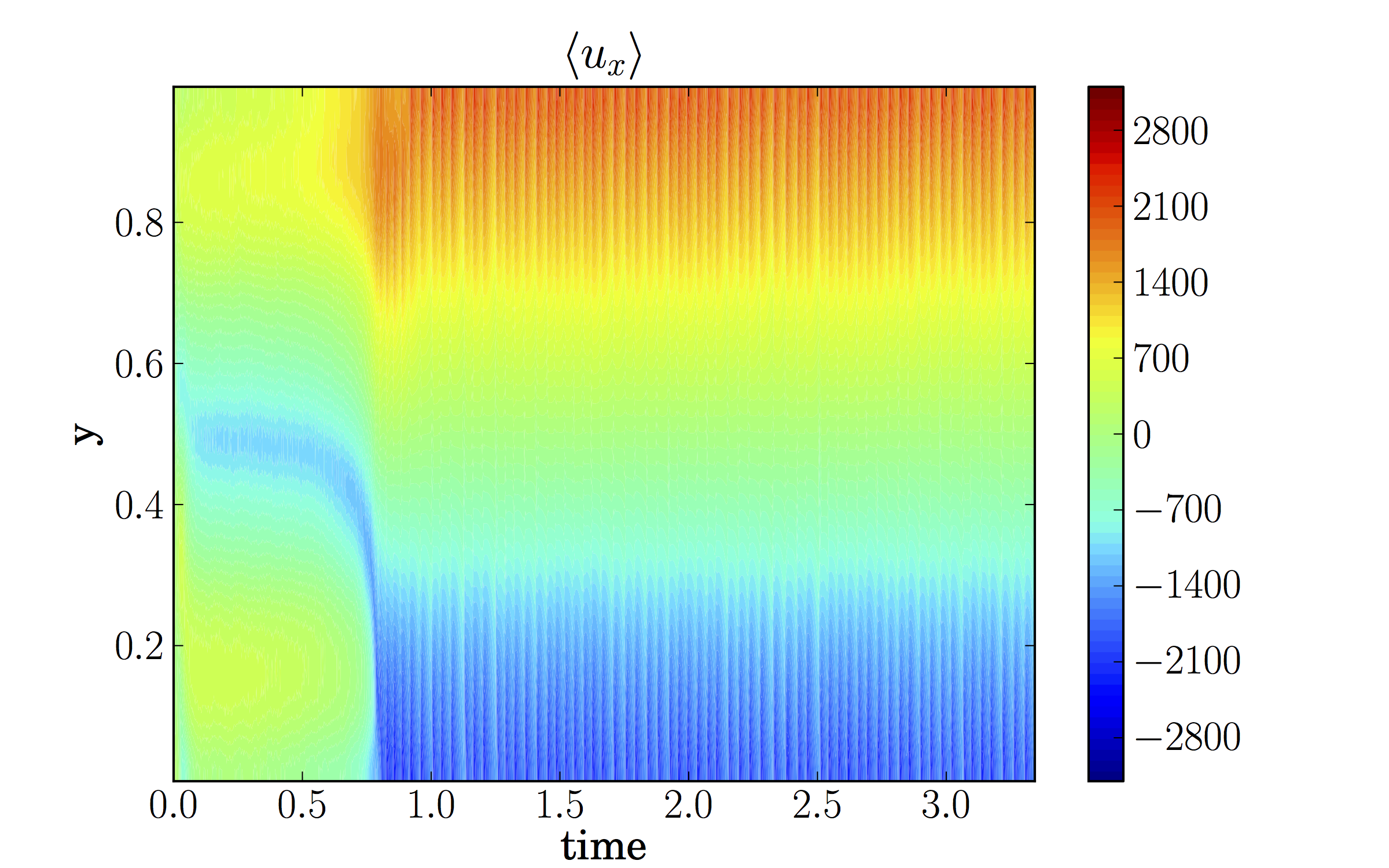}
\caption{Hovmoller plots of DNS solutions for (a) Case A  (b) Case B (c) Case C as defined in Table 1.}
\label{fig_dyn}
\end{figure}

A summary of the dynamics and the parameters for which they are found is given in Table~\ref{table}. Case A is moderately rotating and  the Rayleigh number 
is not high. For this choice of parameters, the convective turbulence interacts with the rotation to produce large-scale jets --- here large-scale 
indicates that the jets are found on the scale of domain whilst the convective cells take a smaller scale. Snapshots of the temperature perturbuation $\theta$ and the horizontal velocity $u_x$ in the saturated regime  shows both the jet structure and the mechanism for the formation of the jet see Figure~\ref{fig_dyn_snap}(a,b). Convective cells, in this case stretching across the domain, interact with the rotation and tilt away from the vertical. This non-trivial systematic tilt leads to the formation of non-zero Reynolds stresses and hence to zonal flows (see \cite{bh1993,jra2003}). In this case the zonal flows are prograde at the top of the domain and retrograde at the bottom. However, owing to the Boussinesq up-down symmetry of the system, solutions with the other parity can be found for different initial conditions. Similar behaviour leading to the generation of mean flows can be found for regular rotating convection in a plane layer in the presence of a non-aligned gravity and rotation vector \cite{ct2016}.
Increase of the rotation rate (and corresponding increase in $Ra$) as in Case B, leads to convection cells being driven on much smaller lengthscales (see Figure~\ref{fig_dyn_snap}(c)). In turn the Reynolds stresses are modified so that multiple jets are found. In this case 7 jets are formed. Although these these are now on a scale smaller than the computational domain, they are still on a scale larger than that for the convective cells, as shown in figure~\ref{fig_dyn_snap}(c,d). In both Case A and B the dynamics reaches a statistically steady state where the mean flows are quasi-steady as shown in the Hovmoller plots in Figure~\ref{fig_dyn}(a,b). These demonstrate that  the jets driven by the Reynolds stresses are fixed in position.

\begin{figure}[!h]
\centering\includegraphics[width=2.5in]{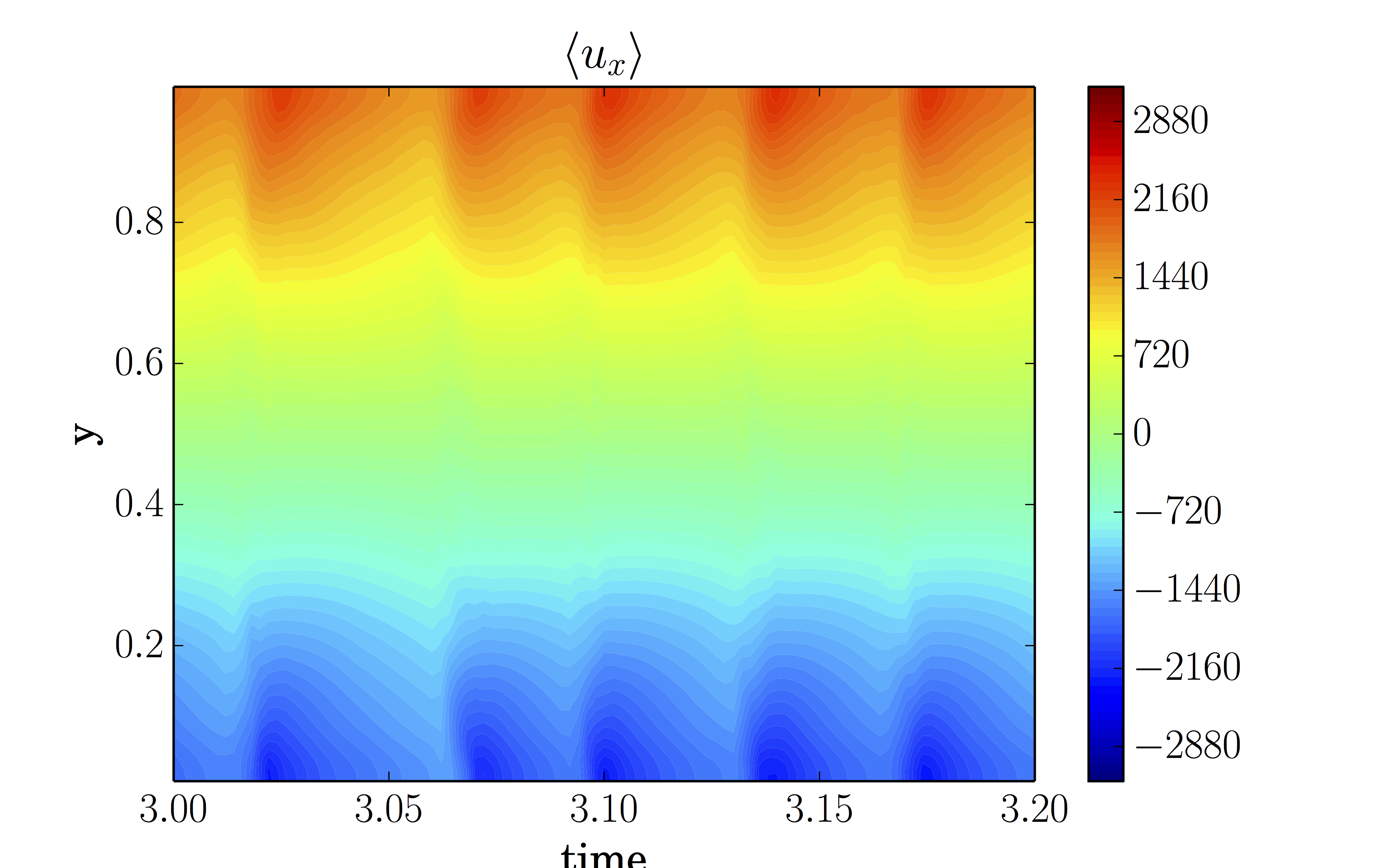}
\includegraphics[width=2.5in]{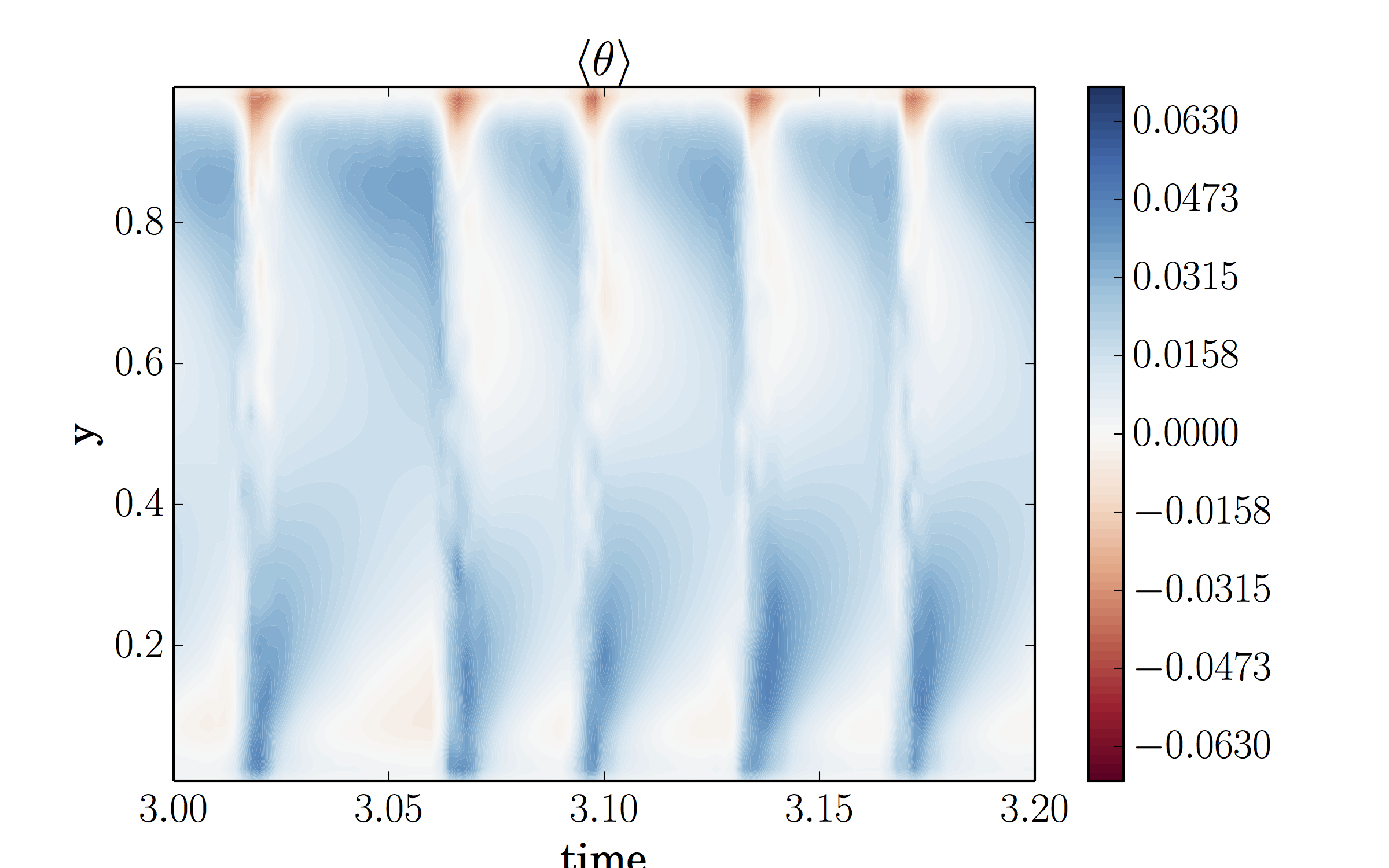}
\caption{Close up Hovmoller plots for (a) $\langle u_x \rangle$ and (b) $\theta$ for DNS of the bursting solution of Case C showing relaxation oscillations.}
\label{fig_clup}
\end{figure}

\begin{figure}[!h]
\centering\includegraphics[width=5.5in]{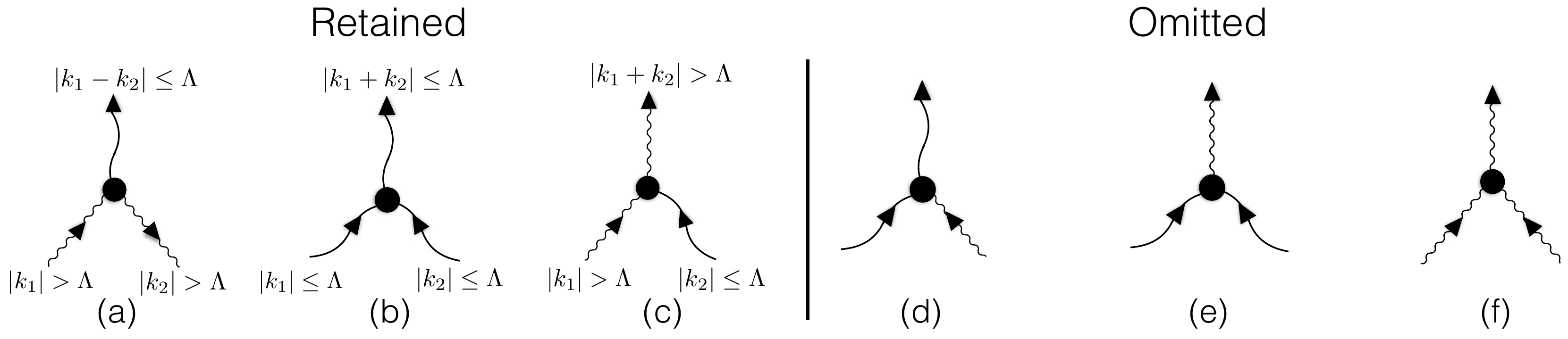}
\caption{Feynman Diagrams showing retained and discarded mode interactions.}
\label{interactions}
\end{figure}

\begin{figure}[!h]
\centering\includegraphics[width=2.5in]{./uxRa76000_all.png}
\includegraphics[width=2.5in]{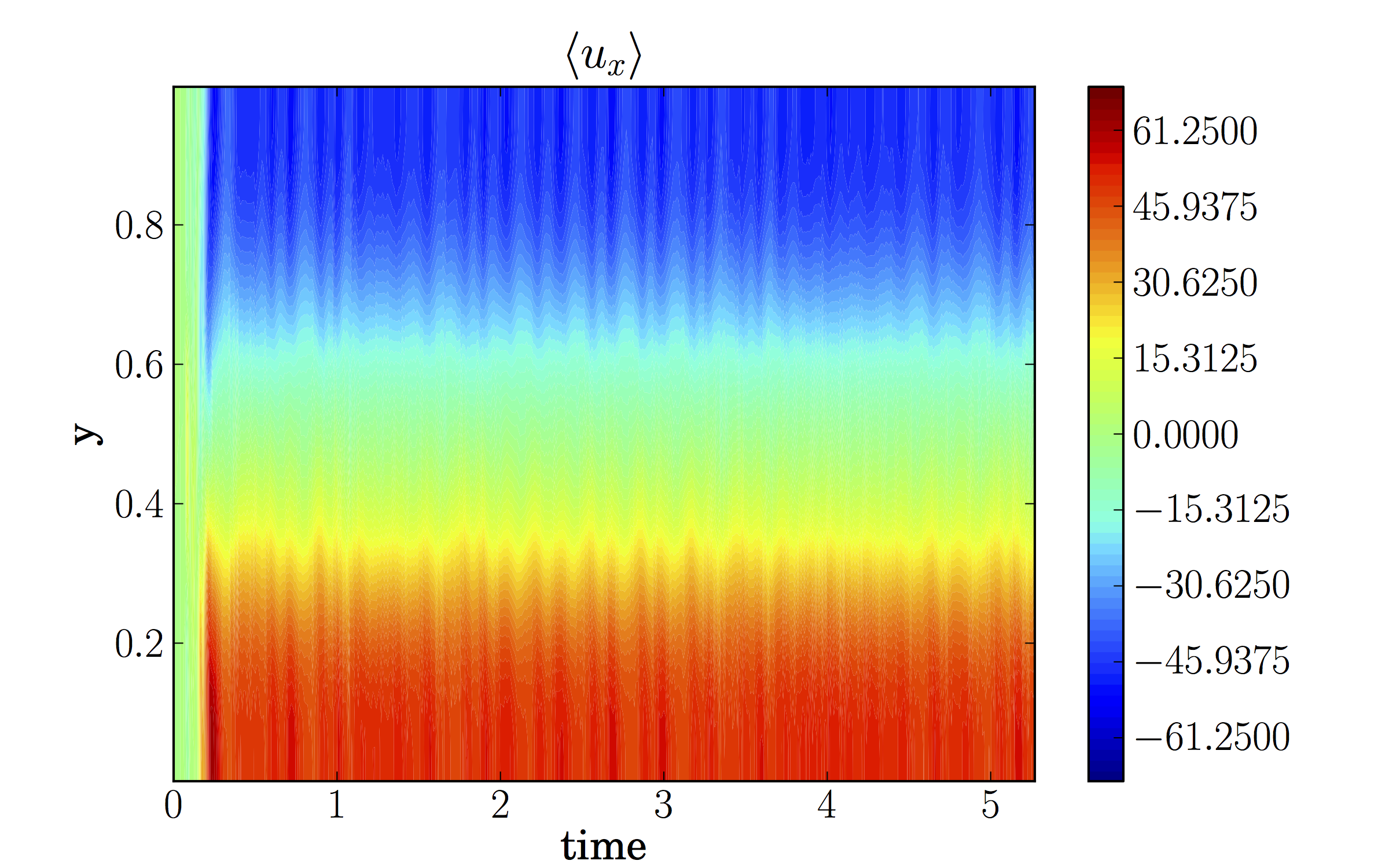}
\includegraphics[width=2.5in]{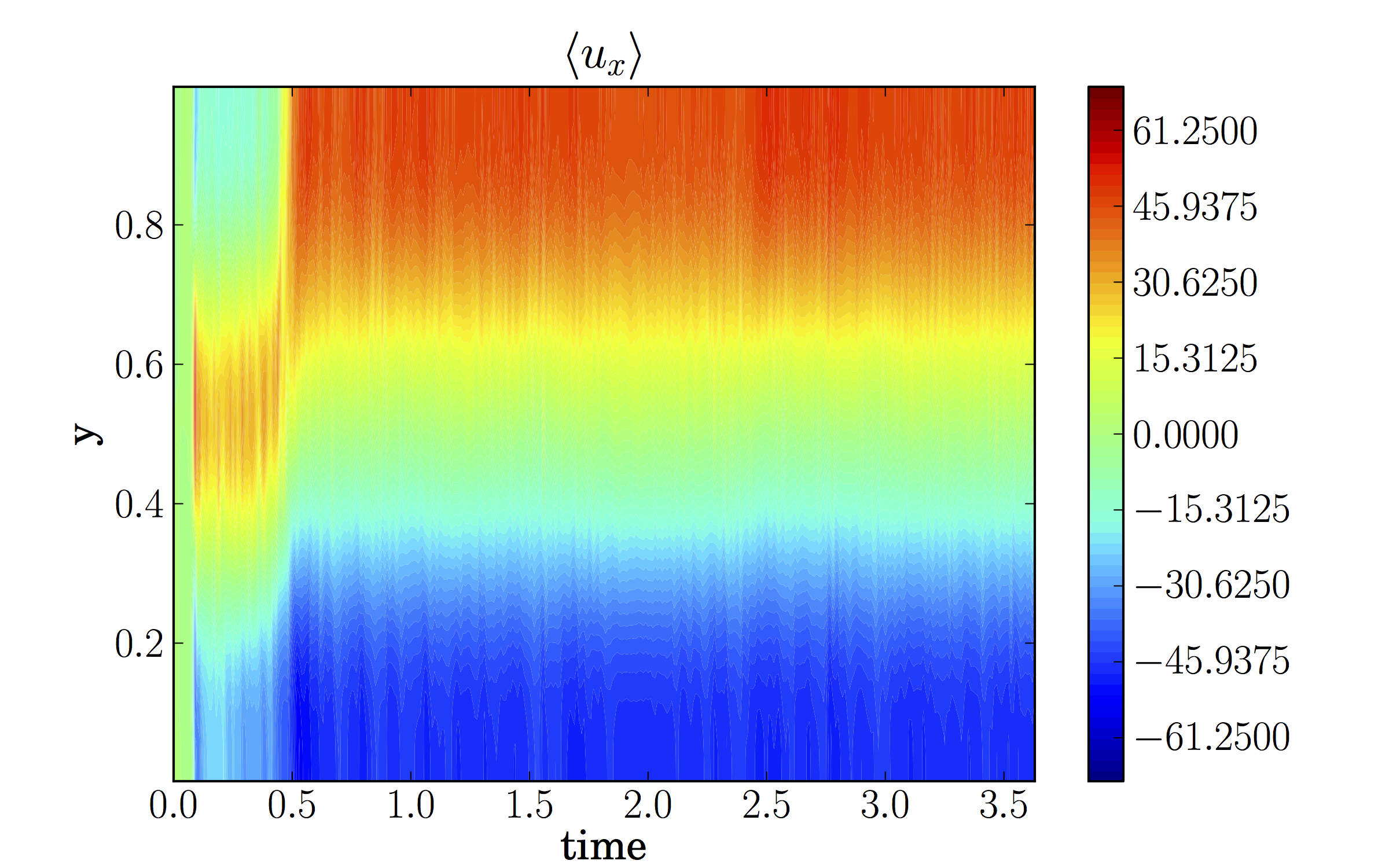}
\includegraphics[width=2.5in]{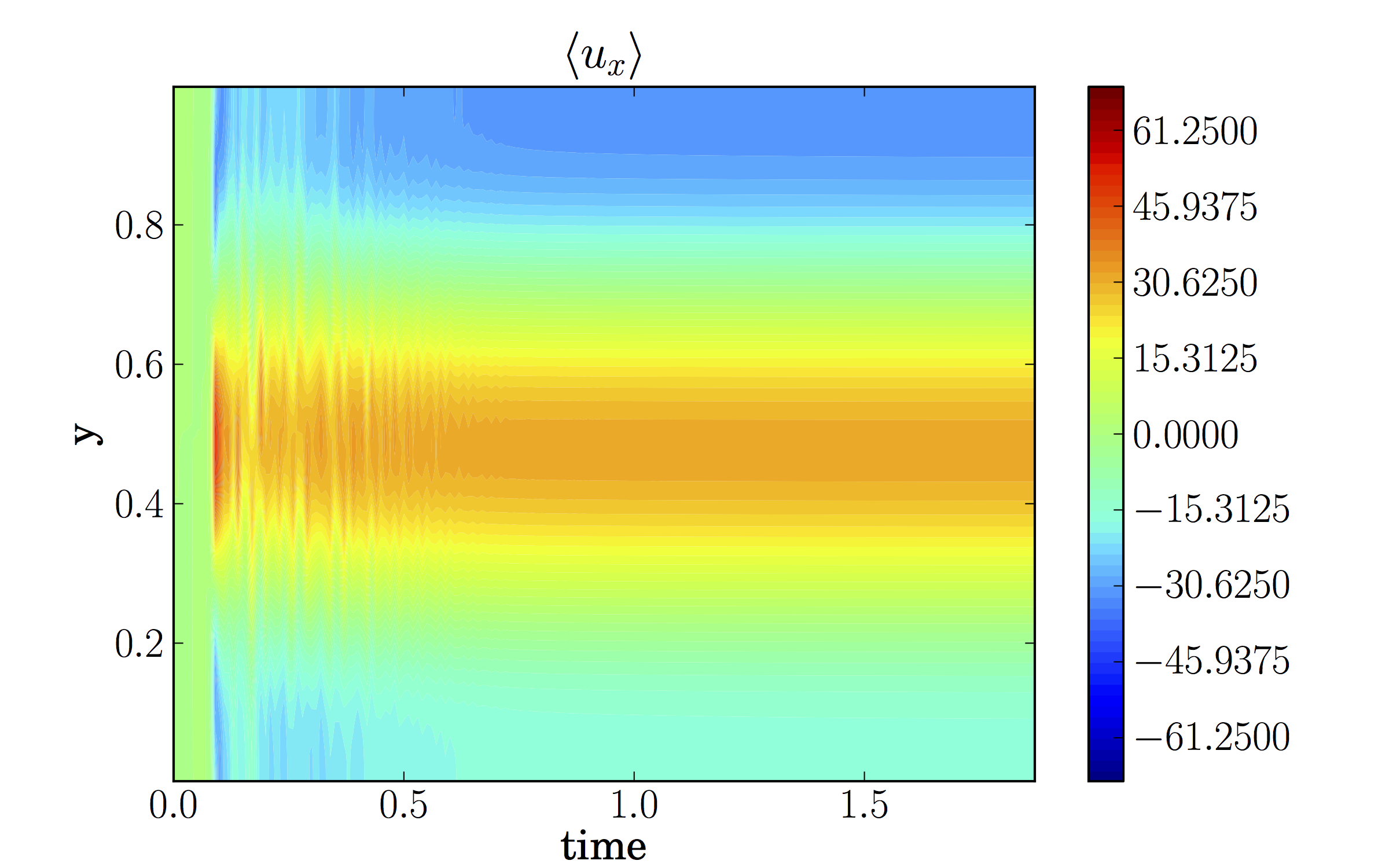}
\caption{Hovmoller plots for Case A for (a) DNS (b) GQL with $\Lambda=5$ (c) GQL with $\Lambda=1$ and (d) QL.}
\label{fig_hov_ls}
\end{figure}

\begin{figure}[!h]
\centering\includegraphics[width=2.5in]{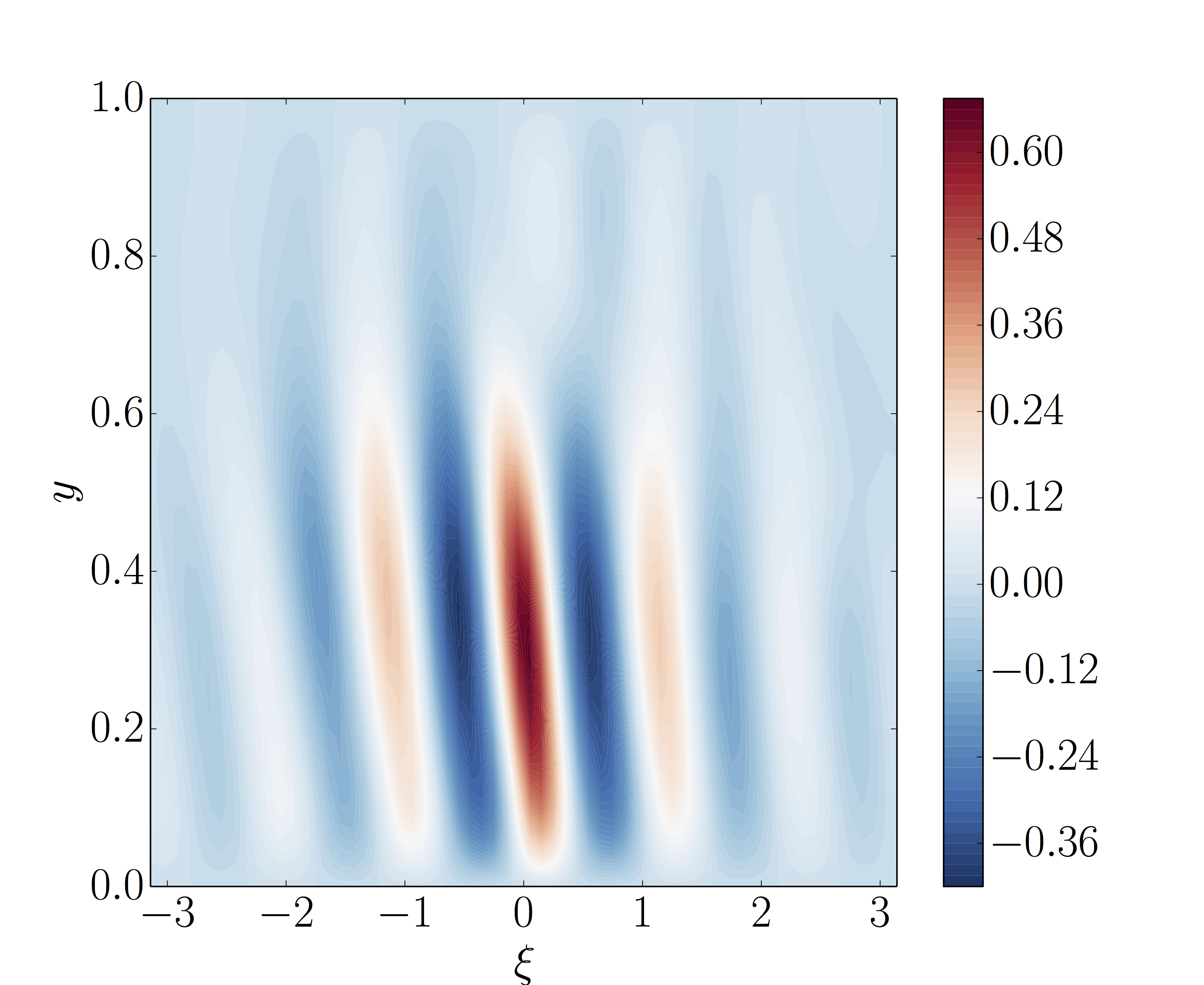}
\includegraphics[width=2.5in]{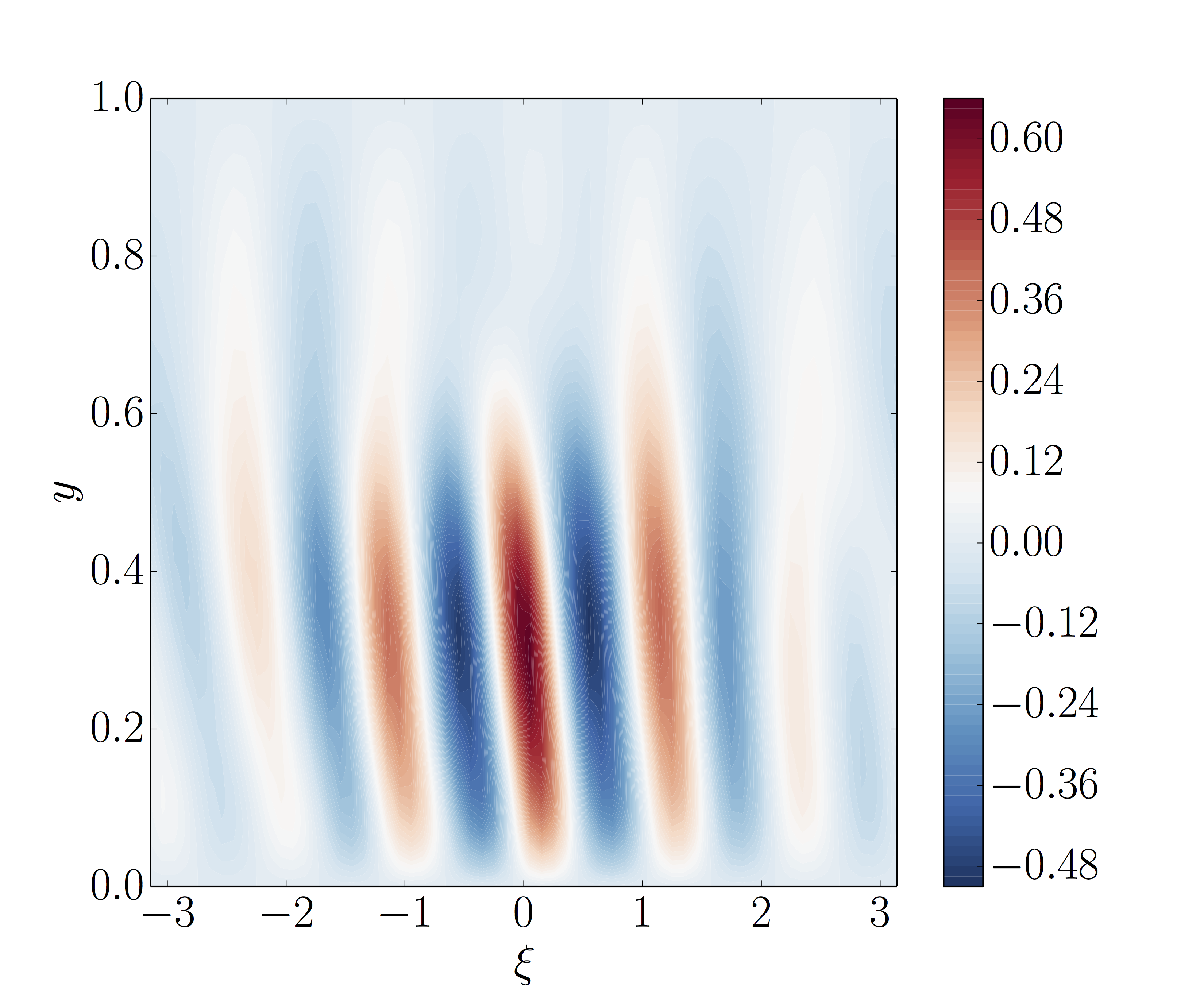}
\includegraphics[width=2.5in]{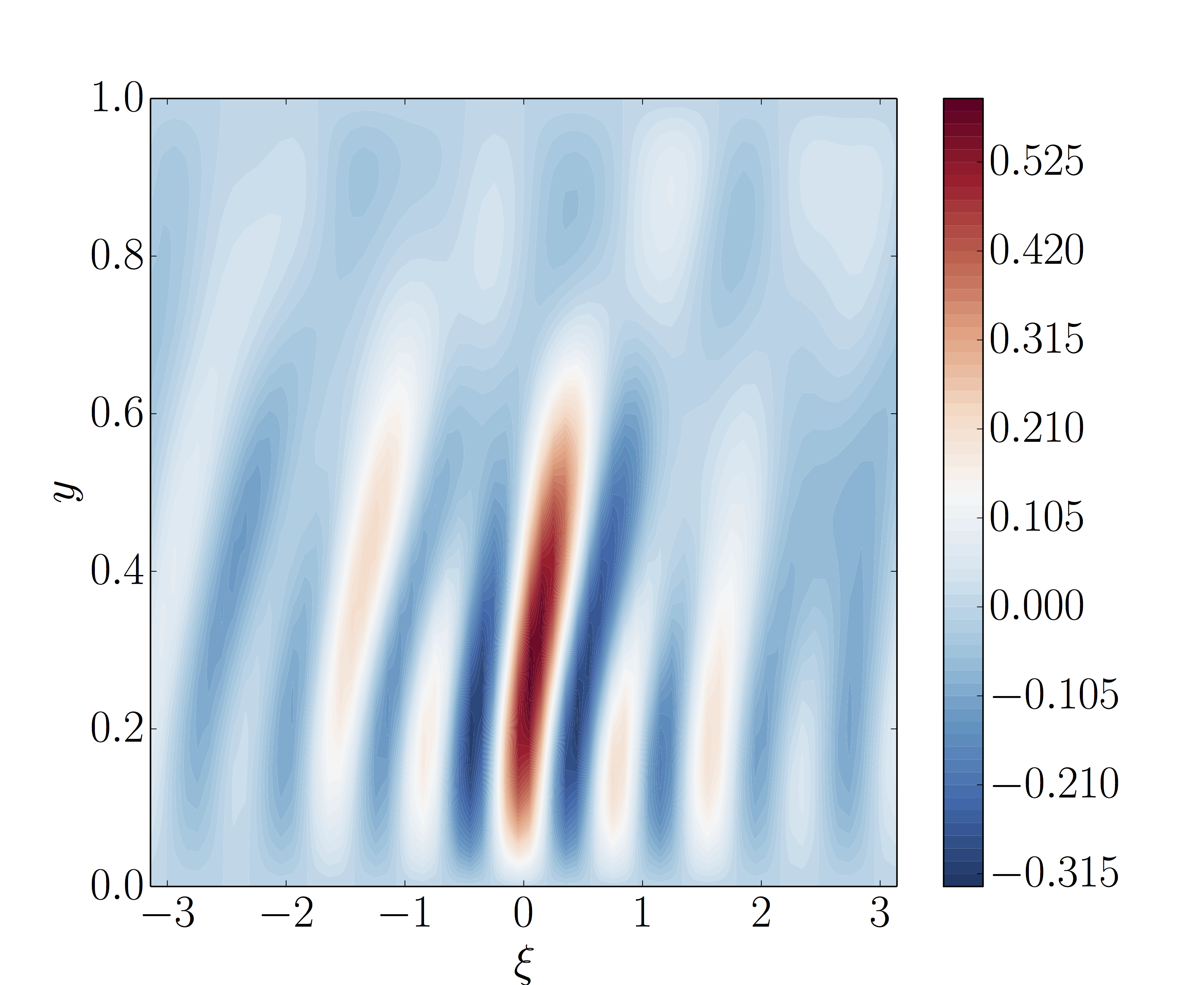}
\includegraphics[width=2.5in]{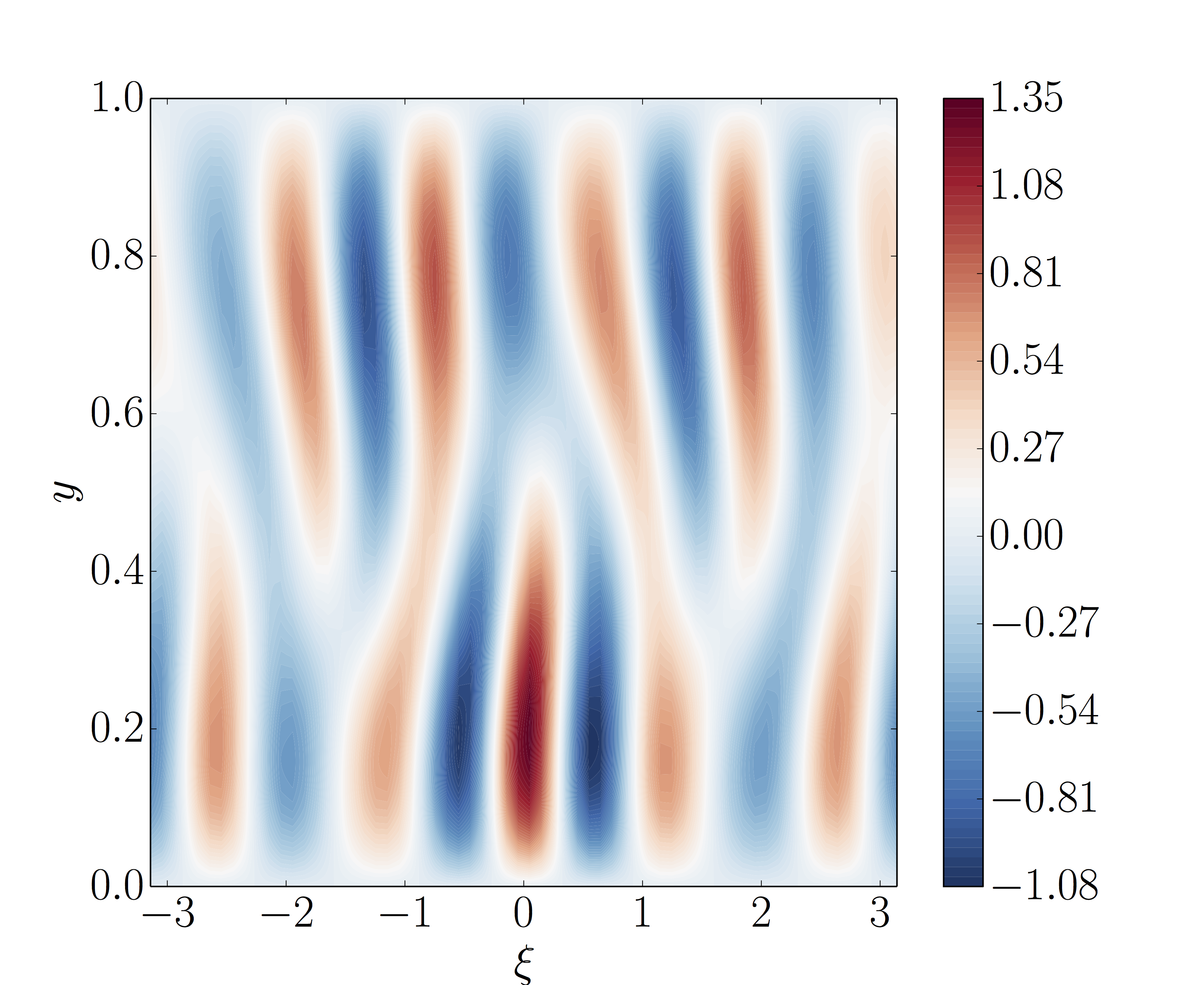}
\caption{Second Cumulants $c_{\theta \theta}$ for Case A for (a) DNS 
(b) GQL with $\Lambda=5$ (c) GQL with $\Lambda=1$ and (d) QL.}
\label{fig_sc_76}
\end{figure}

\begin{figure}[!h]
\centering\includegraphics[width=2.5in]{./ux_7e5_all.png}
\includegraphics[width=2.5in]{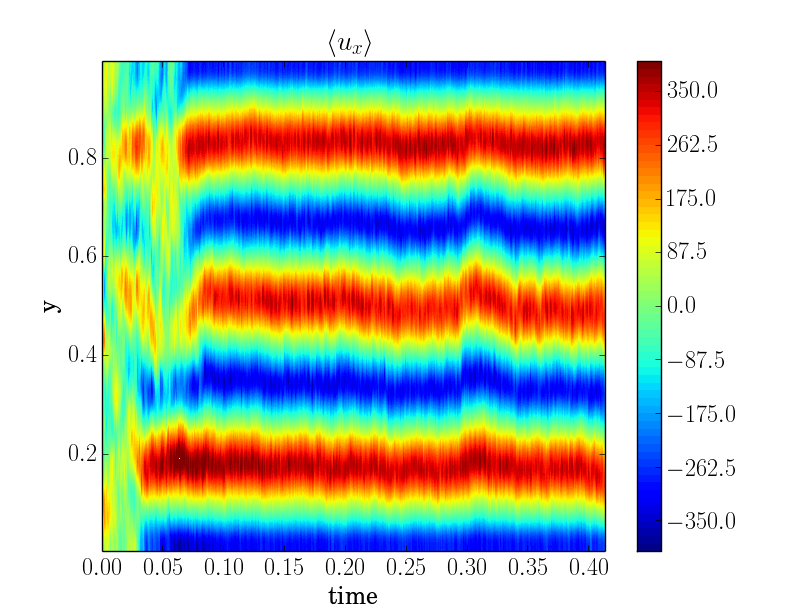}
\includegraphics[width=2.5in]{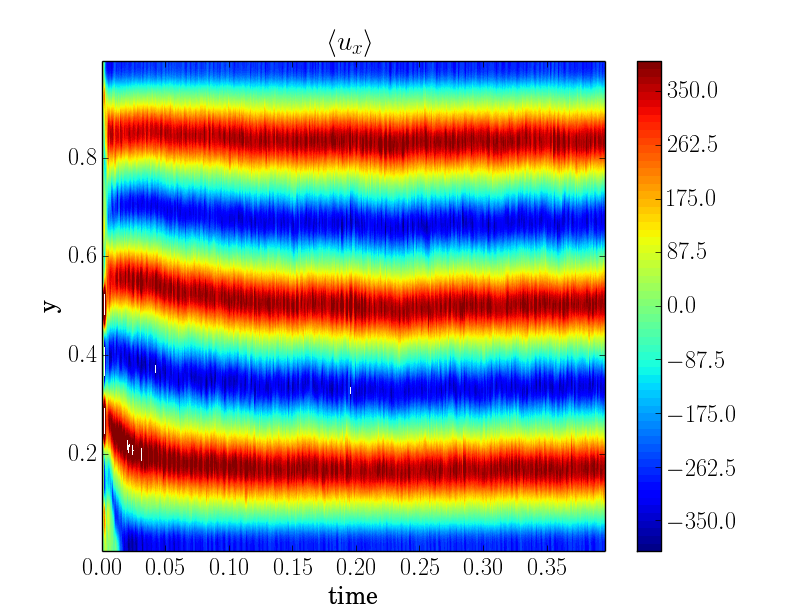}
\includegraphics[width=2.5in]{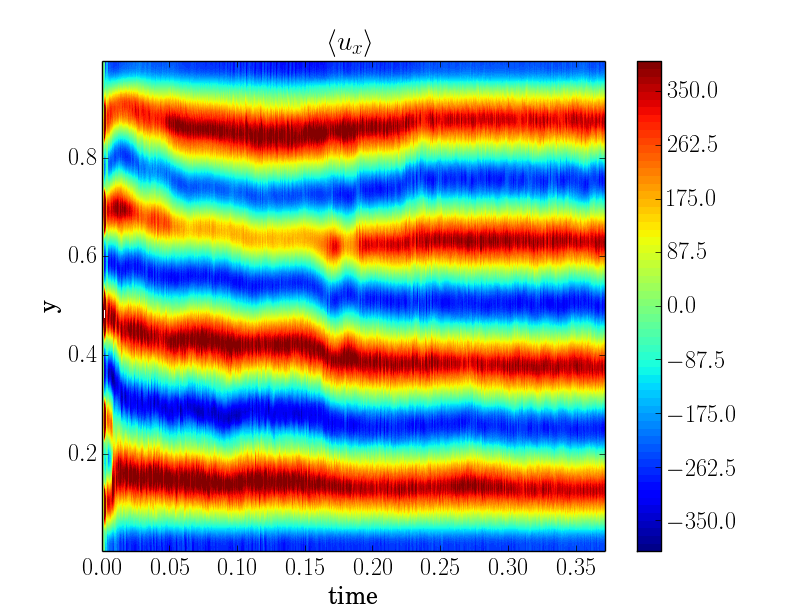}
\caption{Hovmoller plots for Case B for (a) DNS (b) GQL with $\Lambda=5$ (c) GQL with $\Lambda=1$ and (d) QL.}
\label{fig_hov_ss}
\end{figure}

\begin{figure}[!h]
\centering\includegraphics[width=2.5in]{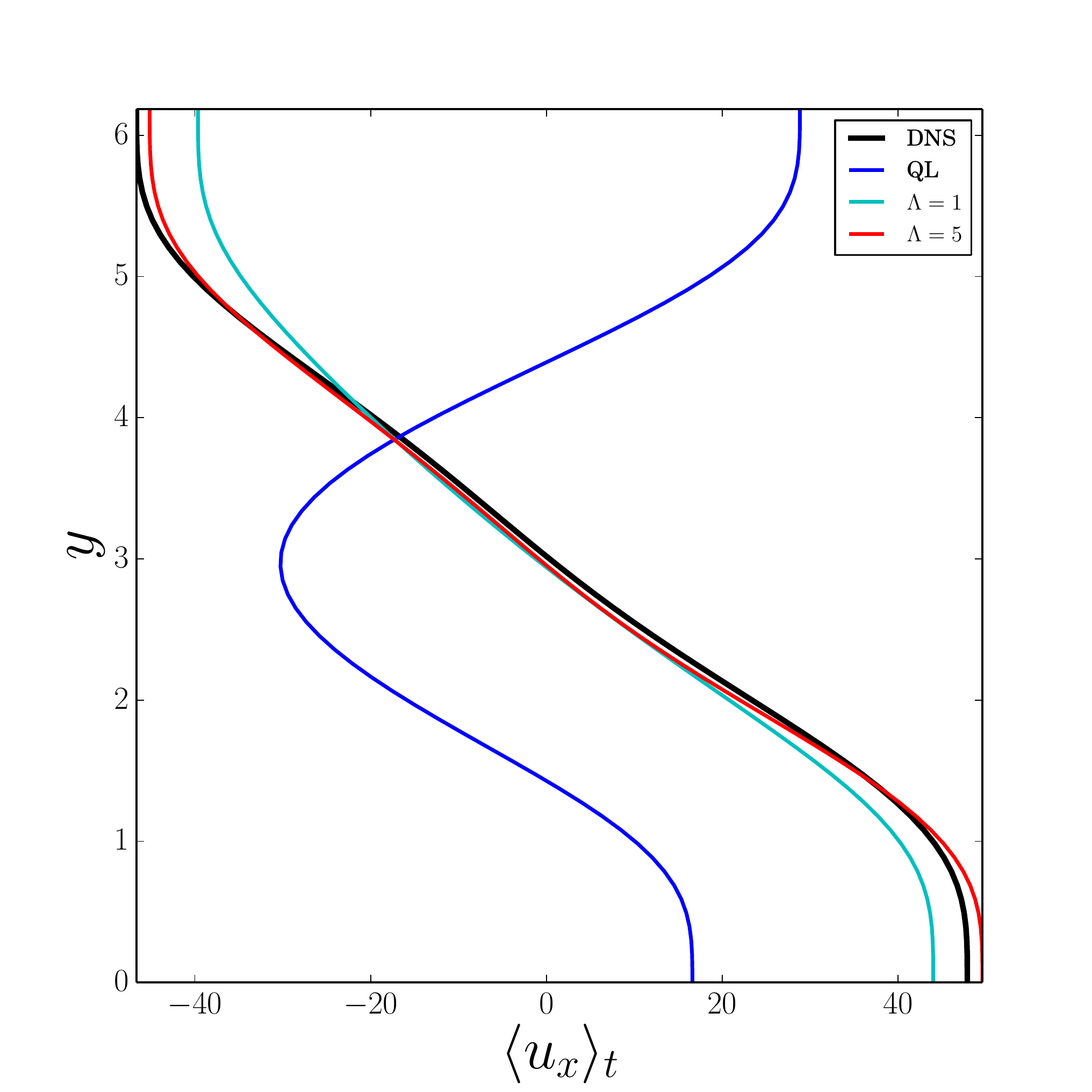}
\includegraphics[width=2.5in]{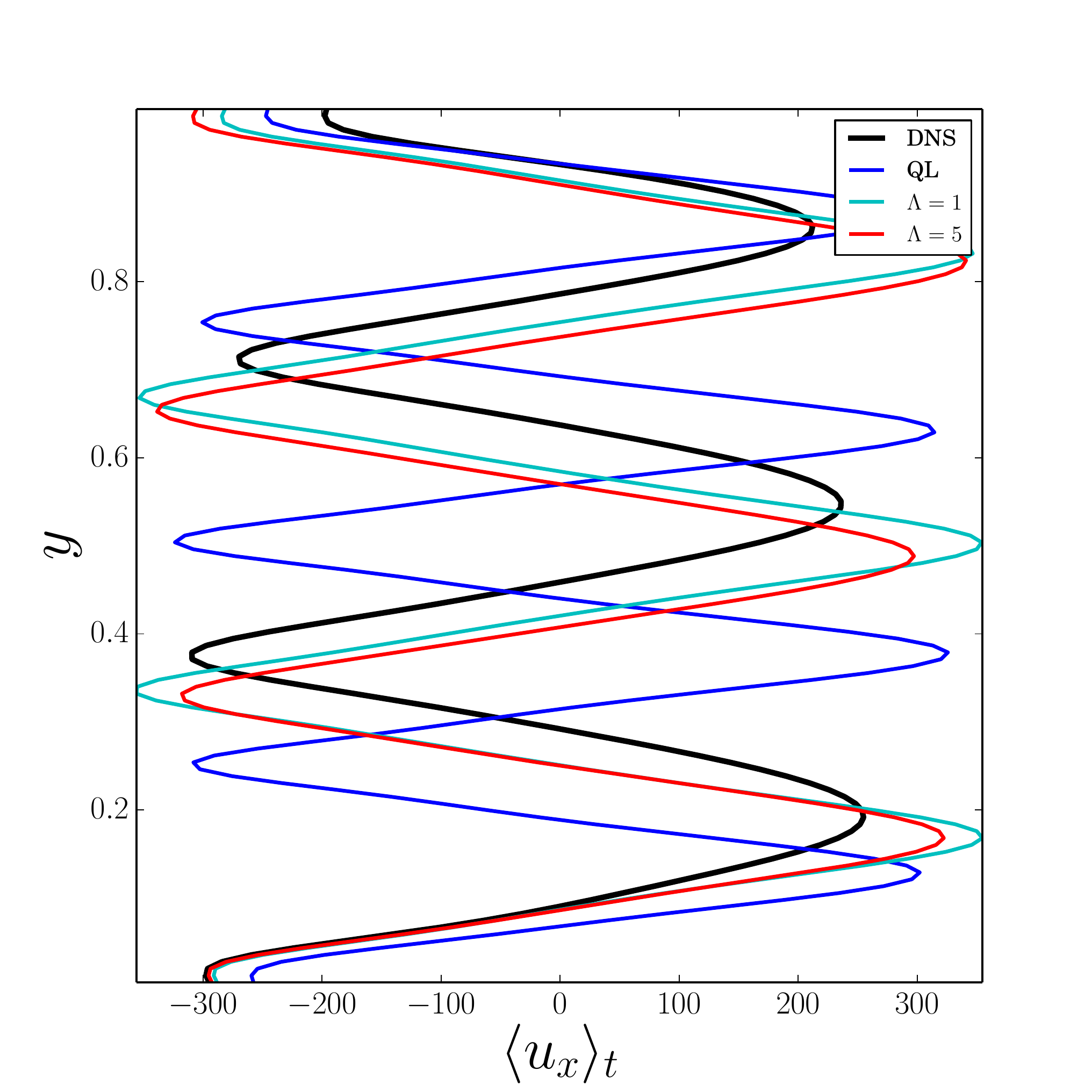}
\caption{Comparison of mean flows averaged over last third of evolution for (a) Case A (b) Case B.}
\label{fig_av}
\end{figure}

\begin{figure}[!h]
\centering\includegraphics[width=2.5in]{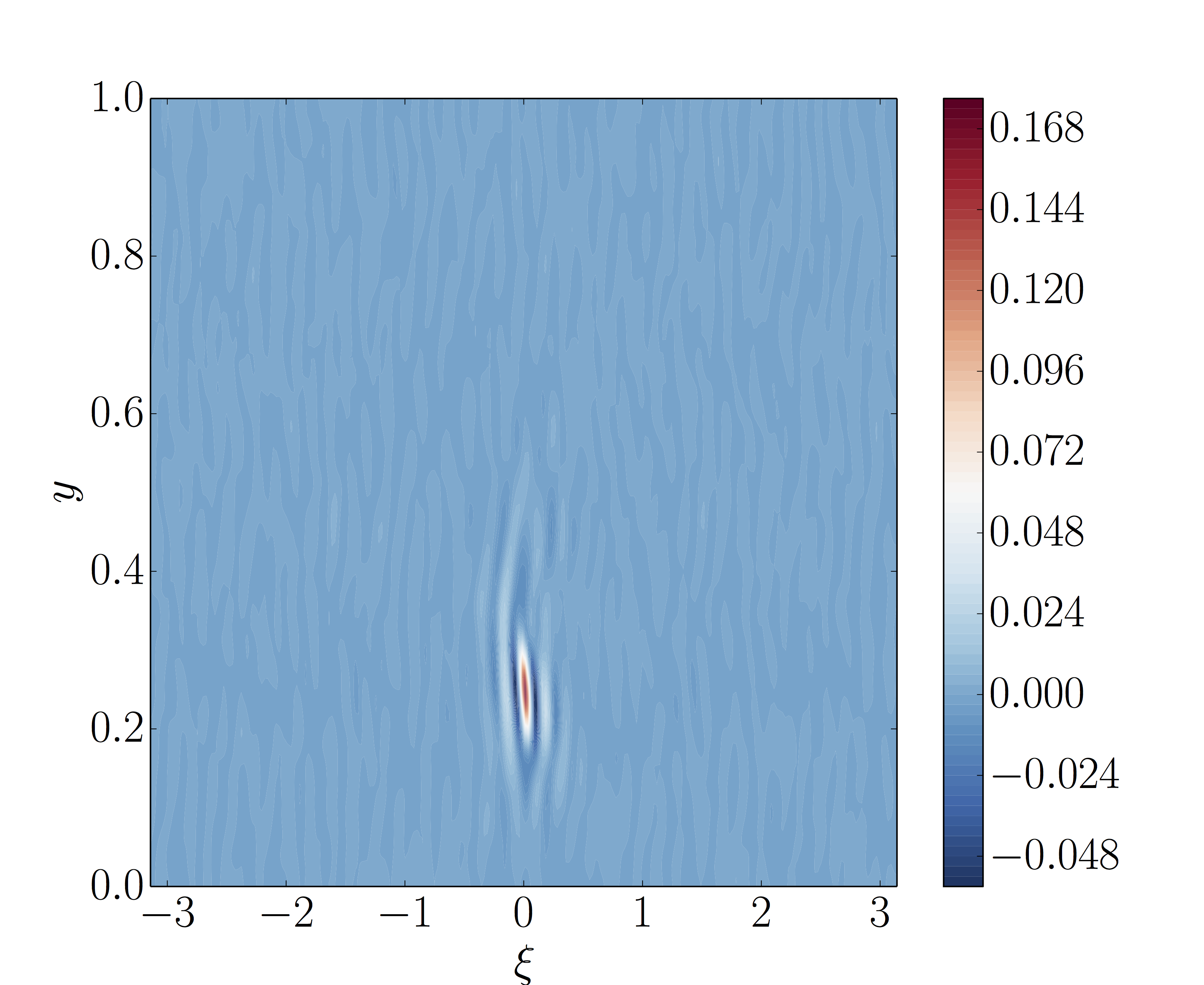}
\includegraphics[width=2.5in]{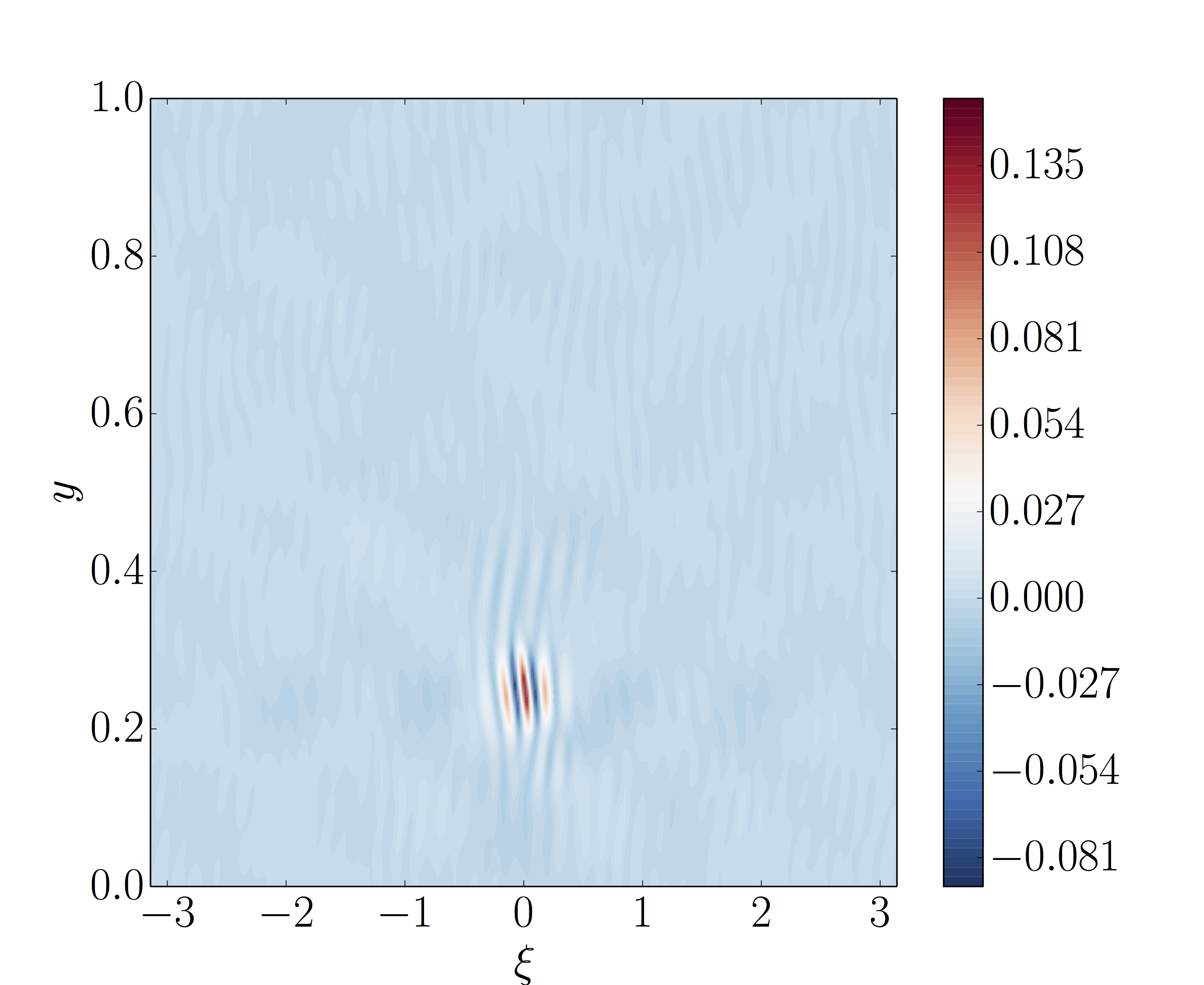}
\includegraphics[width=2.5in]{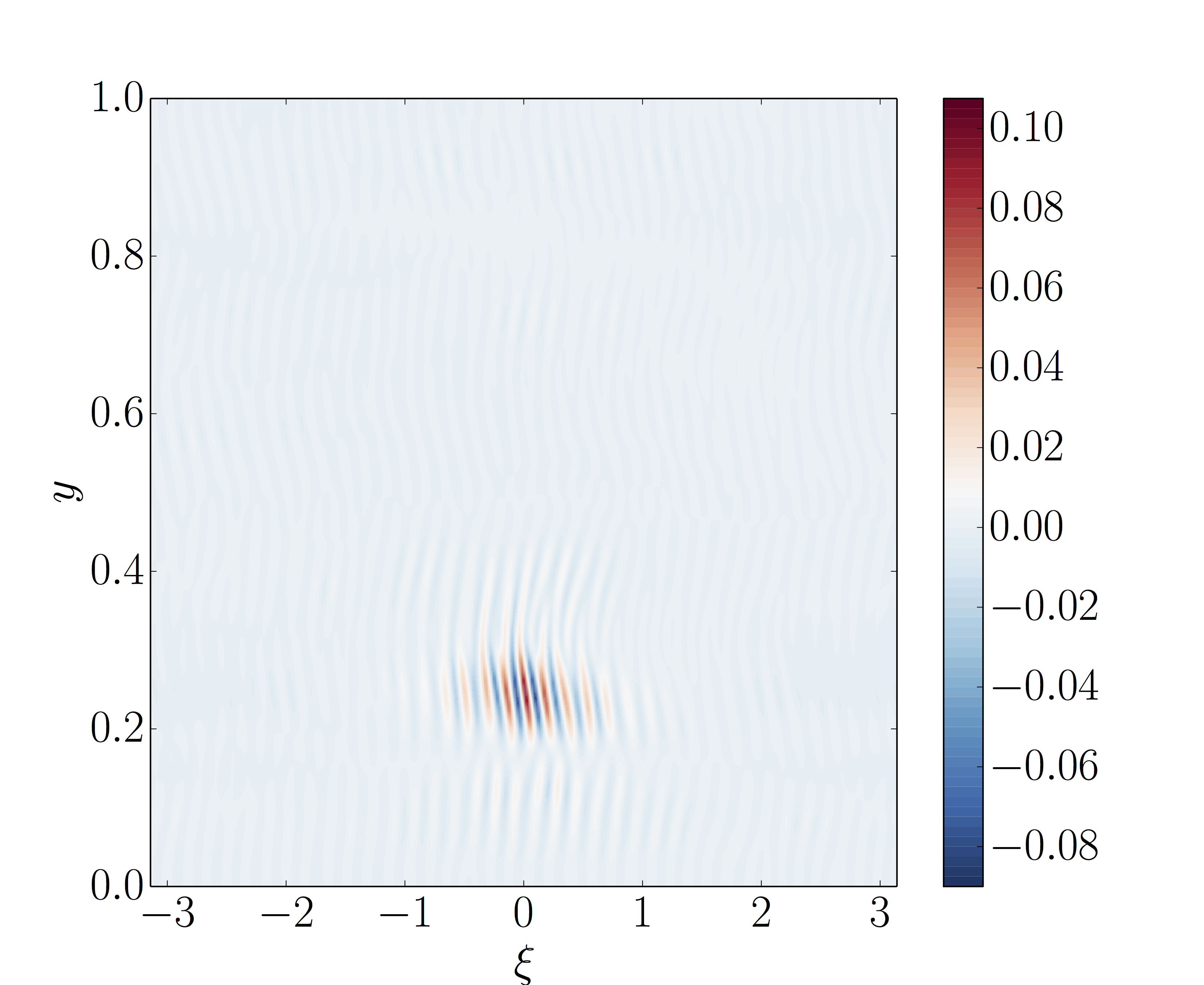}
\includegraphics[width=2.5in]{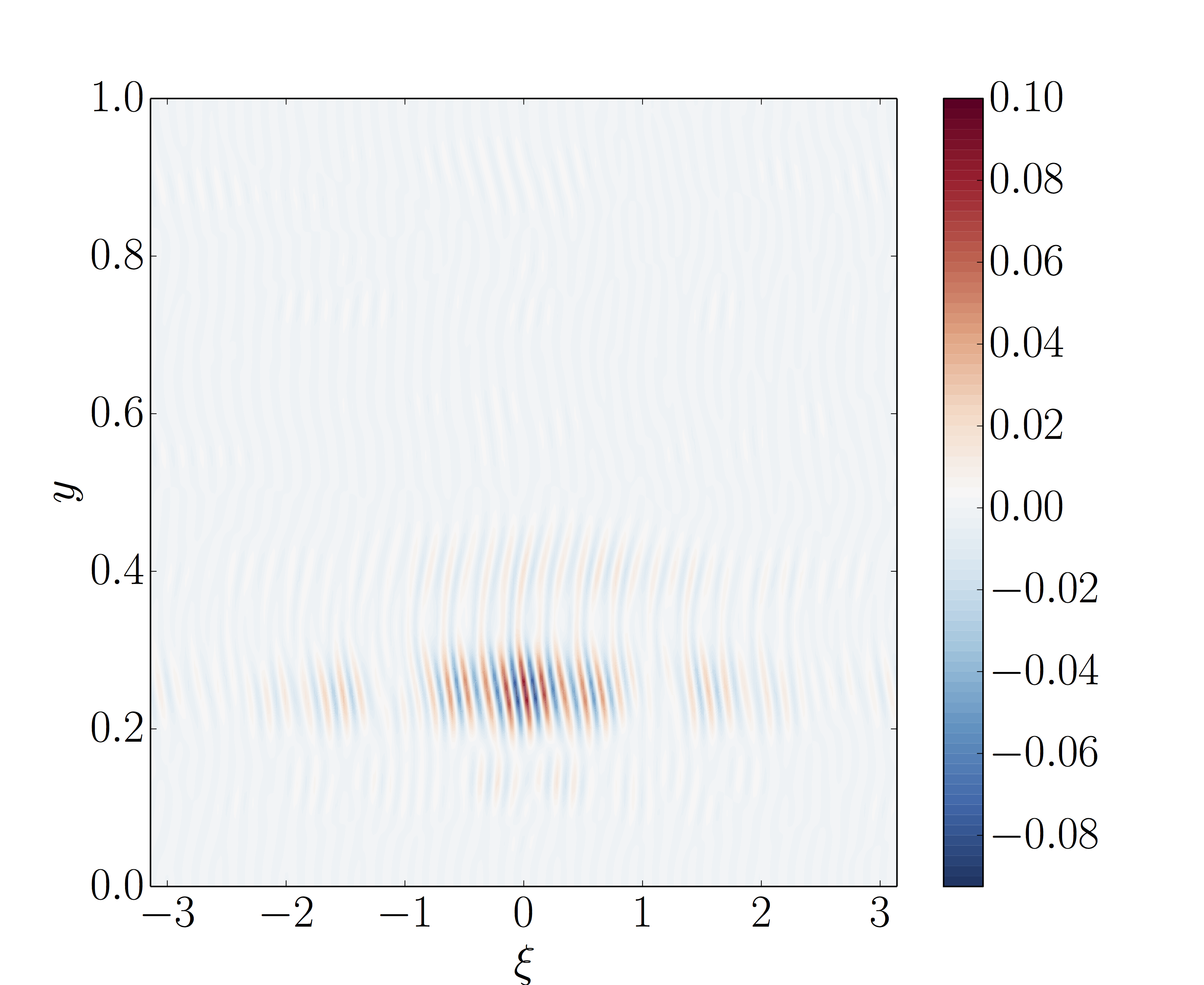}
\caption{Second Cumulants $c_{\theta \theta}$ for Case B for (a) DNS 
(b) GQL with $\Lambda=5$ (c) GQL with $\Lambda=1$ and (d) QL.}
\label{fig_sc_ss}
\end{figure}

\begin{figure}[!h]
\centering\includegraphics[width=2.75in]{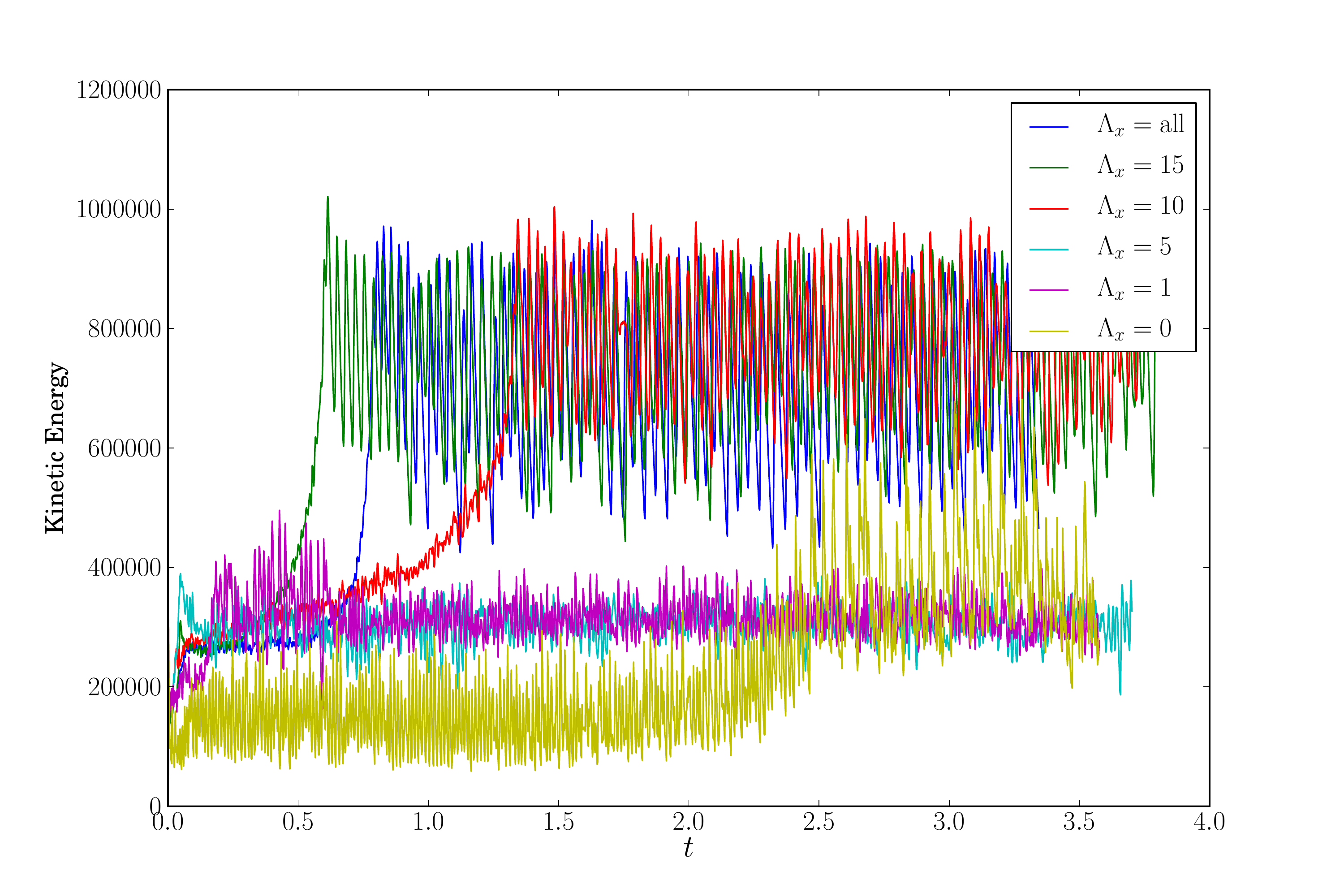}
\includegraphics[width=2.75in]{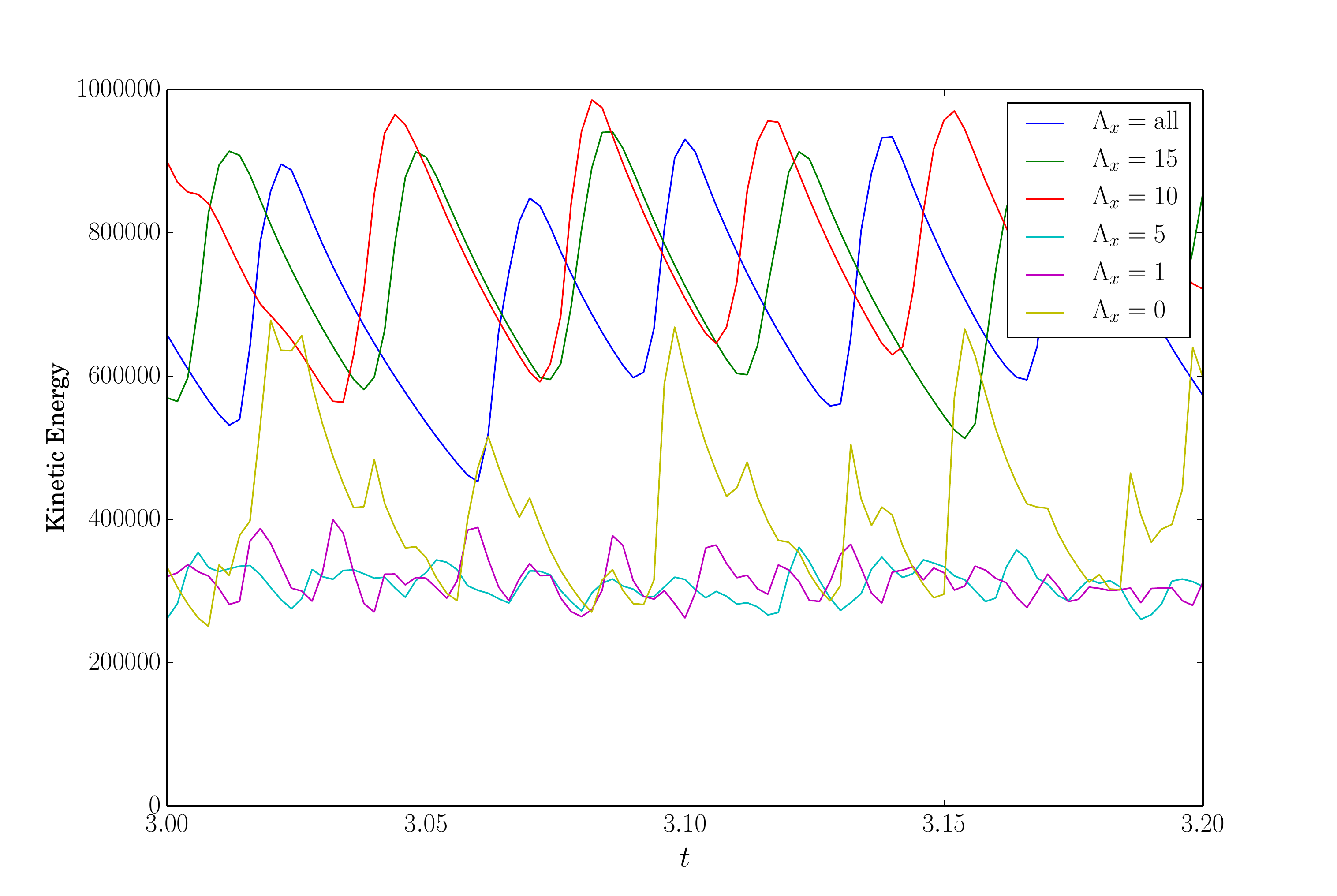}
\caption{Timeseries of kinetic energy density for Case C (a) full timeseries (b) zoom of timeseries showing relaxation oscillations.}
\label{fig_ke_burst}
\end{figure}

\begin{figure}[!h]
\centering\includegraphics[width=2.5in]{./ux5e5Ra8e8_all.png}
\includegraphics[width=2.5in]{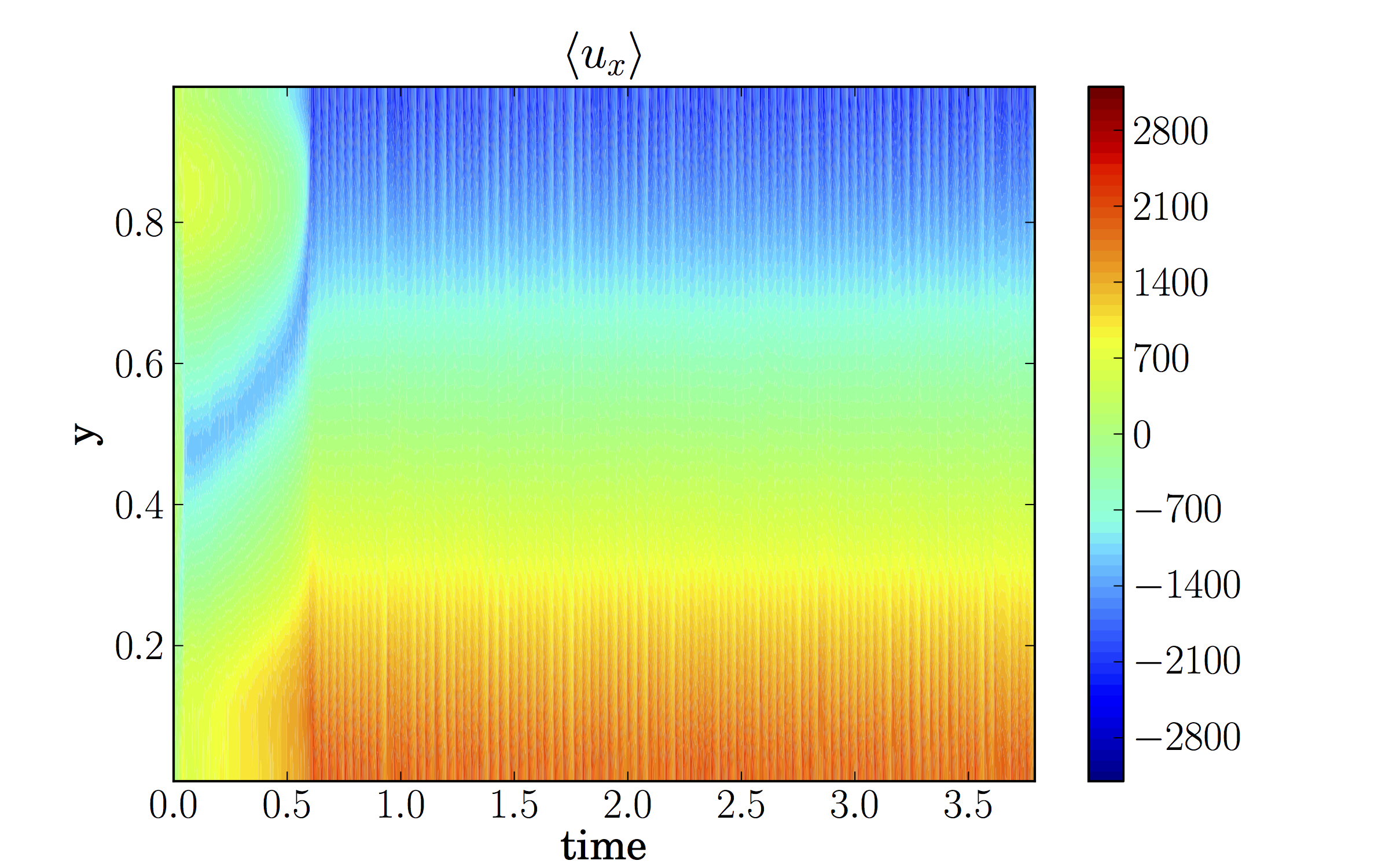}
\includegraphics[width=2.5in]{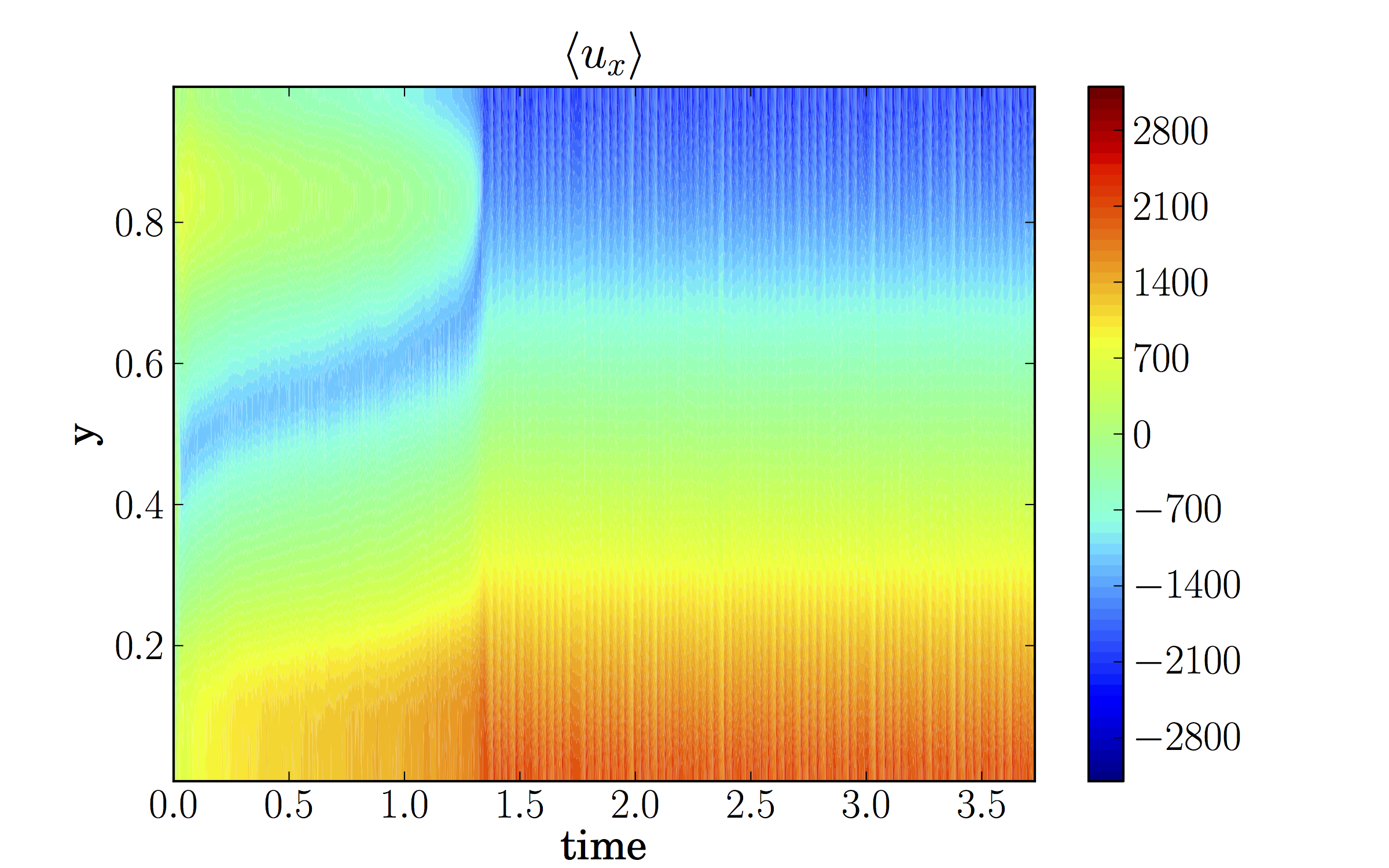}
\includegraphics[width=2.5in]{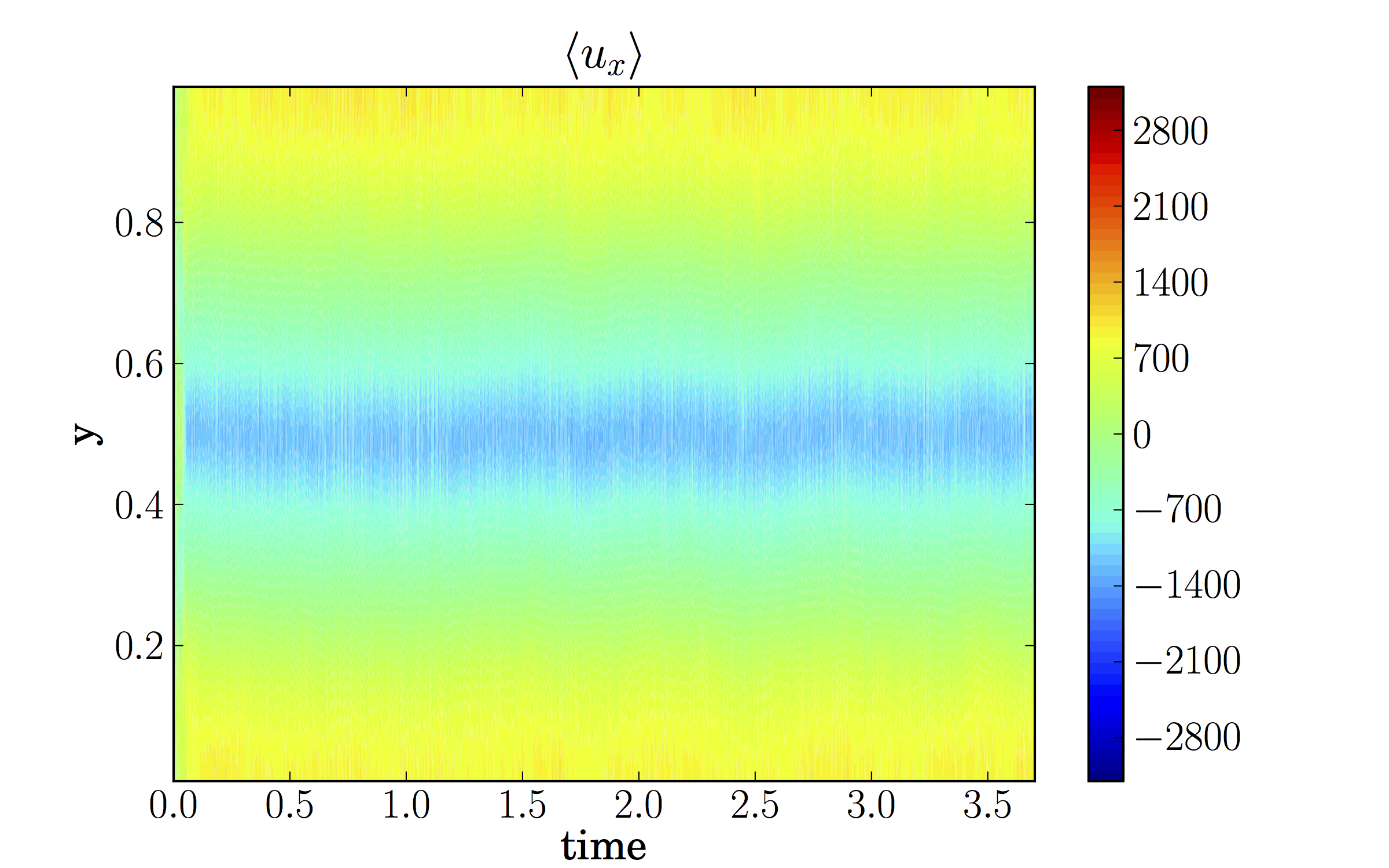}
\includegraphics[width=2.5in]{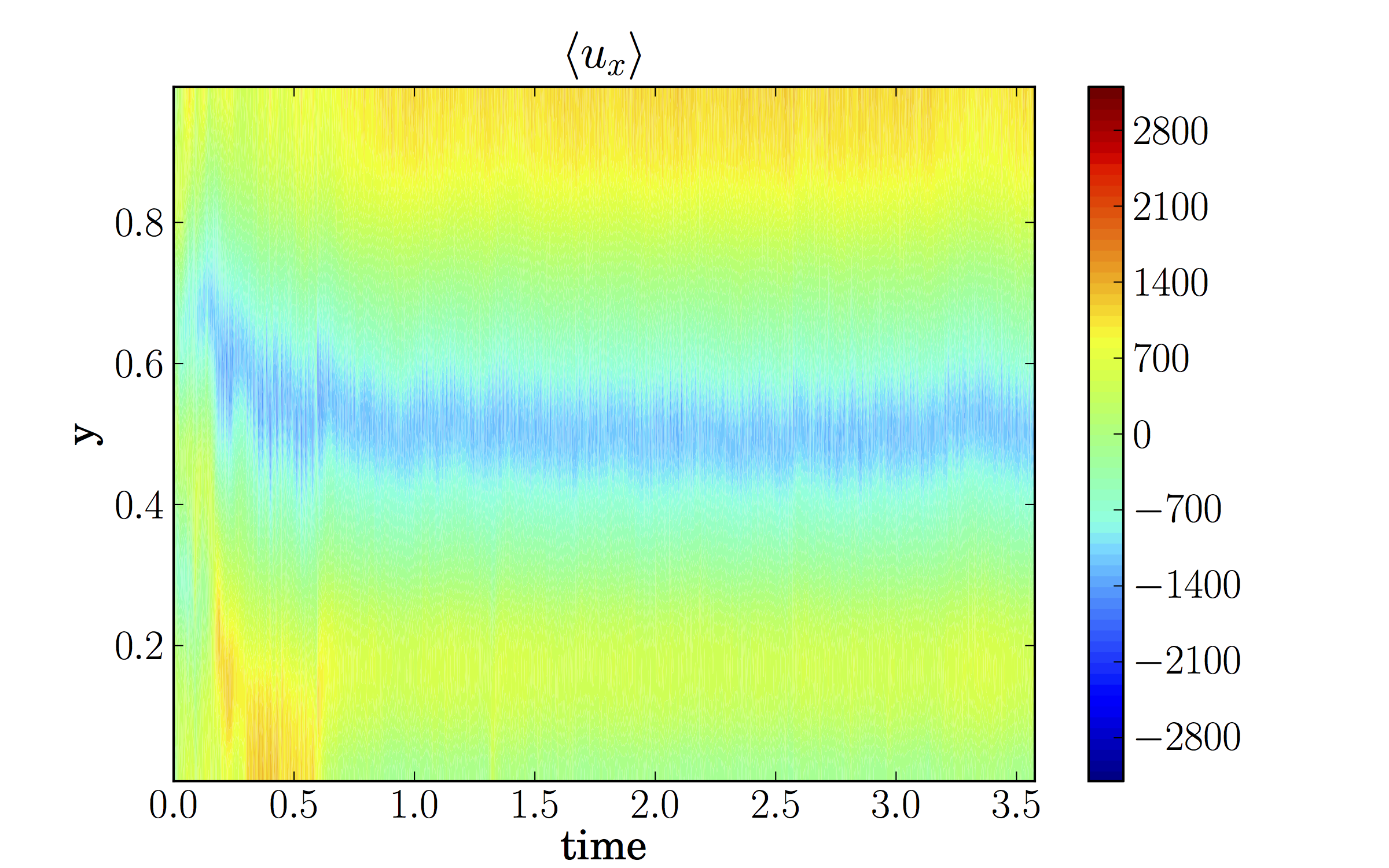}
\includegraphics[width=2.5in]{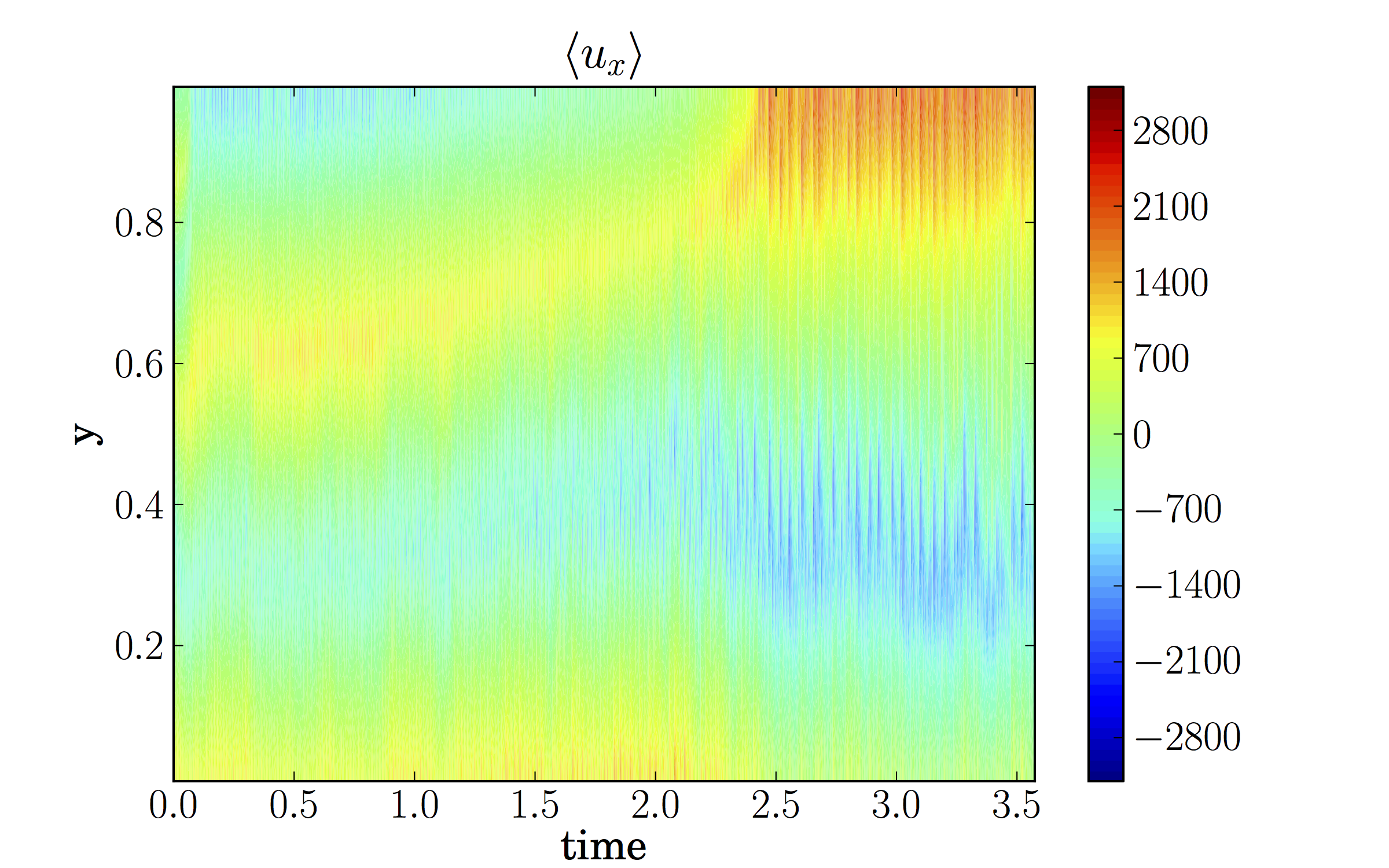}
\caption{Hovmoller plots for Case C for (a) DNS (b) GQL with $\Lambda=15$ (c) GQL with $\Lambda=10$
(d) GQL with $\Lambda=5$ (e) GQL with $\Lambda=1$ and (f) QL.}
\label{fig_hov_burst}
\end{figure}

Figures~\ref{fig_dyn}(c) show the dynamics of the large-scale jets for Case C via a long time-scale Hovmoller plot. Here the jets although large-scale exhibit non-trivial temporal
variation. Up until $t \approx 1$ the large-scale flow has three quasi-steady jets; however this behaviour is transient and the solution eventually takes the form of two jets. Interestingly at this point the system exhibits bursting between states in the form of a relaxation oscillator. A typical cycle of the oscillation proceeds as follows (as shown in Figure~\ref{fig_clup}). 
Convection interacts with the rotation to drive a mean shear (as before). However in this case the shear is strong enough to act as a barrier to transport and hence suppress the convection (which is itself the source of the shear). The shear therefore slowly decays and the transport barrier disappears. The cycle is then free to proceed again. This predator-prey type dynamics between the mean flows and the turbulence has been seen before in a number of systems involving zonal flows;  these include models of convection, the interaction of zonal flows with turbulence in tokamak plasmas, dynamo models \cite{t1997,kgd2015,bs2006} and the magnetorotational instability \cite{OML2011}. Figure~\ref{fig_clup} show snapshots of $u_x$ and $\theta$ in both the 
weak and strong convection phases.

The dynamics of the annulus system is therefore rich and varied. Complicated interactions between mean flows, turbulence and driving lead to non-trivial dynamics and mean flows and
so this system represents a formidable challenge for approximations such as QL and GQL. It is the efficacy of these approximations that we shall investigate in the next section.


\subsection{Evaluation of the QL and GQL approximations}
\label{GQL}


In order to construct the QL and GQL systems we perform an idempotent decomposition of the streamfunction and temperature into large-scale and small-scale modes, i.e.\ we set, say, $\psi(x, y)= \psi_l + \psi_h$, where
\begin{equation}
\psi_l(x, y)
=\sum_{k=-\Lambda}^{\Lambda}{\psi}_{k}(y) \,e^{i  {2 \pi k x}/{L_x}}, \quad \quad
\psi_h = \psi - \psi_l, 
\end{equation}
where $\psi_l$ and $\psi_h$ are the `low' and `high' wavenumber modes respectively.   Similarly ${\hat \theta}(x,y) = {\hat \theta}_l+{\hat \theta}_h$.

Having made this decomposition we derive equations for the evolution of the low and high modes, making use of the Generalised Quasilinear Approximation. That is we set retain the nonlinear interaction between low modes and low modes and that between high modes and high modes   to yield low modes in the low mode equation and also retain the interaction between low modes and high modes to yield high modes in the high mode evolution equation. All other nonlinear interactions are discarded (as shown in Figure~\ref{interactions}. As this ensures that triad interactions are removed in pairs, this process is an example of constrained triad decimation in pairs \cite{kraichnan1985} and is therefore guaranteed to conserve quadratic invariants in the dissipationless system. Crucially we perform this decimation for both the inertial nonlinearity in the Navier-Stokes equation {\it and} the advective nonlinearity in the temperature equation. Hence we set
\begin{eqnarray}
\dfrac{\partial \nabla^2 {\psi}_l}{\partial t} + J({\psi}_l, \nabla^2 {\psi}_l) + J({\psi}_h,\nabla^2 {\psi}_h)- \beta \dfrac{\partial {\psi}_l}{\partial x} &=& -\dfrac{Ra}{Pr} \, \dfrac{\partial {\hat \theta}_l}{\partial x} - C |\beta|^{1/2} \nabla^2 {\psi}_l + \nabla^2 \nabla^2 {\psi}_l, \nonumber\\
\dfrac{\partial \nabla^2 {\psi}_h}{\partial t} + J({\psi}_l, \nabla^2 {\psi}_h) +J({\psi}_h, \nabla^2 {\psi}_l) - \beta \dfrac{\partial {\psi}_h}{\partial x} &=& -\dfrac{Ra}{Pr} \, \dfrac{\partial {\hat \theta}_h}{\partial x} - C |\beta|^{1/2} \nabla^2 \psi_h + \nabla^2 \nabla^2 \psi_h,\nonumber\\
\dfrac{\partial {\hat \theta}_l}{\partial t} + J(\psi_l, {\hat \theta}_l) +  J(\psi_h, {\hat \theta}_h)&=& - \dfrac{\partial {\psi_l}}{\partial x} + \dfrac{1}{Pr} \nabla^2 {\hat \theta}_l,\nonumber\\
\dfrac{\partial {\hat \theta}_h}{\partial t} + J(\psi_l, {\hat \theta}_h) + J(\psi_h, {\hat \theta}_l) &=& - \dfrac{\partial {\psi}_h}{\partial x} + \dfrac{1}{Pr} \nabla^2 {\hat \theta}_h.
\label{psiTeqns_gql}
 \end{eqnarray} 
The QL system can therefore be recovered by setting $\Lambda=0$ (so that only the $k_x=0$ mode is counted as a low mode), whilst DNS is reproduced by setting the cut-off
$\Lambda$ to be the highest wavenumber of the spectral truncation of DNS. Intermediate values of $\Lambda$ yields the GQL approximation.

We start by considering how well QL and GQL perform for the steady large-scale jets of Case A. Figure~\ref{fig_hov_ls} shows a comparison of the Hovmoller plots for $\langle u_x \rangle$ as a function of $y$ and $t$ for the cases of DNS, $\Lambda=5$, $\Lambda=1$ and QL ($\Lambda=0$). These indicate that GQL performs well even at $\Lambda=5$ and $\Lambda=1$, with both of these settling down into a two-jet solution (note that the sign of the jet is not important owing to the Boussinesq symmetry). However QL performs not so well, selecting a weaker three-jet solution. This is confirmed in the time averages of $\langle u_x \rangle$ (denoted $\langle u_x \rangle_t$) shown in Figure~\ref{fig_av}(a). Here the time average is taken over the last third of the evolution. Clearly GQL (even at a cutoff $\Lambda=1$) is performing well in representing the first cumulant (or zonal mean). 

Accurately describing the second cumulants represents more of a test for the QL and GQL approximation. Figure~\ref{fig_sc_76} shows,  the second cumulant for temperature $c_{\theta \theta}$ defined as
\begin{equation}
c_{\theta \theta}(\xi,y_1,y_2) = \int \theta'(x_1,y_1) \theta'(x_1+\xi,y_2) dx_1,
\end{equation}
where $\xi = x_2 - x_1$. The cumulant is shown, averaged over the last third of the temporal evolution,  for a representative choice of $y_1 = 0.25$ as a function of $\xi$ and $y_2=y$, for DNS, QL and GQL (at $\Lambda=1$, $5$).
For DNS the second cumulant, as expected peaks around the reference point $(\xi,y)=(0,0.25)$ and with a lengthscale of variation given by the width of a typical convective structure and a height of half the domain. The structure of the second cumulant also displays a characteristic tilt, which leads to the non-trivial Reynolds stress and is also shaped by the shear --- note that the sense of the tilt is reversed if the shear is reversed Far from the reference point the correlations die away and hence the second cumulant tends to zero. GQL (at both $\Lambda=1$, $5$) does a good job of reproducing this structure, but QL overstates the importance of long-range correlations, as it overrepresents the importance of thermal Rossby waves in the system. This type of behaviour has been seen before for QL systems \cite{mct2016,chmt2016,tm2017}.

Figure~\ref{fig_hov_ss} compares the evolution of the mean flows for Case B for the various approximations with that of DNS. All the approximations are able to capture the driving of multiple jets by the convection, however the QL approximation does not reproduce the correct number of jets (having two more). The GQL approximation performs better again, with the correct number of jets of approximately the correct amplitude (as shown in Figure~\ref{fig_av}(b)). The second cumulants in Figure~\ref{fig_sc_ss} show the localised nature of the interactions owing to the reduction in lengthscale of the convection. GQL again provides a good approximation of the correlations. As the number of modes
in the GQL approximation is decreased the approximation becomes unsurprisingly worse, with the amplitude of the local correlation being underestimated and the correlations becoming
delocalised, again because of the overestimation of the stability of thermal Rossby waves.

FInally in this section we consider the bursting case C. For this case, in addition to QL and GQL at $\Lambda=1$ and $\Lambda=5$ we also consider GQL at $\Lambda=10$ and $\Lambda=15$. Figure~\ref{fig_ke_burst}(a) shows the timeseries of the kinetic energy for DNS, GQL (at the various truncations) and QL, with a zoom of the saturated regime shown in Figure~\ref{fig_ke_burst}(b). It is clear that, for this complicated solution, more modes are required in GQL in order for the bursting solution to be accurately represented.
If QL or GQL at $\Lambda=1$ or $\Lambda=5$ is used then the energy is significantly  lower than the ``true'' solution. The reason for this is shown in the Hovmoller plot of
Figure~\ref{fig_hov_burst}. QL and low truncation GQL gets both the amplitude and the number of jets incorrect. Interestingly, although QL exhibits bursting behaviour, low truncation GQL does not (at least for these initial conditions). However by the time $\Lambda=10$, GQL does manage to reproduce faithfully the amplitude, number and bursting
of jets. For such a complicated solution, it is not clear what is the correct procedure  averaging the second cumulants, i.e. whether to average over the whole solution or over bursts and troughs separately so we do not include such a plot in the description of the results.

\section{Conclusions}

We have constructed GQL models for three distinct flow regimes in a Busse annulus model. For each, GQL accurately reproduces key features from the DNS results that QL calculations do not. In particular, for the steady cases (case A and B), GQL with a single extra retained mode ($\Lambda = 1$) correctly predicts the number and amplitude of jets and compares favorably with the DNS for the second cumulant. With $\Lambda = 5$, the GQL second cumulant matches very well to the DNS. For Case C, the significantly more complex bursting dynamics can also be reproduced using GQL, though $\Lambda \ge 10$ in order to do so. The $\Lambda = 10$ case reproduces the amplitude, jet structure, and time-dependence of the relaxation oscillations.

These flows demonstrate that GQL significantly outperforms QL in three dynamically different, far-from-equilibrium flows. In each case, a small number of retained low modes are sufficient to approximate the flow. We have also demonstrated that the rich dynamics offered by Busse annulus flows are particularly challenging to standard QL techniques. Given that closures at second order (e.g. CE2) represent an efficient trade-off between dimensionality and performance, we take the present results to bolster previous results \cite{mct2016,chmt2016,tm2017} that point toward the development of GCE2 techniques as a particularly promising path toward efficient, accurate numerical models of complex geophysical and astrophysical flows.  

\enlargethispage{20pt}

\ethics{This work only involves data obtained from computer simulations.}

\dataccess{The computer code that solved the GQL equations is available as supplementary material.}

\aucontribute{All authors contributed equally to the design, implementation and analysis of the numerical experiments.  All authors gave final approval for publication.}

\competing{We have no competing interests.}

\funding{Insert funding text here.}

\ack{Insert acknowledgment text here.}



\bibliography{paper}
\end{document}